\documentclass[aps,prb,twocolumn,showpacs,superscriptaddress,preprintnumbers,amsmath,amssymb,longbibliography]{revtex4-1}

\usepackage[utf8]{inputenc}
\usepackage{dcolumn}
\usepackage{bm}
\usepackage{ulem}    
\usepackage{xcolor}
\usepackage{gensymb}   
\usepackage{tikz}
\usepackage[pdftex,colorlinks=true,pdfstartview=FitV,linkcolor=blue,citecolor=blue,urlcolor=blue]{hyperref}
\begin{document}
\title{Tuning Elastic Properties of Metallic Nanoparticles by Shape Controlling: \\
From Atomistic to Continuous Models}
\author{Matteo Erbi'}
\affiliation{Laboratoire d'Etude des Microstructures, ONERA-CNRS, UMR104, Universit\'e Paris-Saclay, BP 72, Ch\^atillon Cedex, 92322, France}
\author{Hakim Amara}
\email{hakim.amara@onera.fr}
\affiliation{Laboratoire d'Etude des Microstructures, ONERA-CNRS, UMR104, Universit\'e Paris-Saclay, BP 72, Ch\^atillon Cedex, 92322, France}
\affiliation{Universit\'e de Paris, Laboratoire Mat\'eriaux et Ph\'enom\`enes Quantiques (MPQ), F-75013, Paris, France}
\author{Riccardo Gatti}
\email{riccardo.gatti@onera.fr}
\affiliation{Laboratoire d'Etude des Microstructures, ONERA-CNRS, UMR104, Universit\'e Paris-Saclay, BP 72, Ch\^atillon Cedex, 92322, France}
\date{\today}

\begin{abstract}
 
Understanding and mastering the mechanical properties of metallic nanoparticles is crucial for their use in a wide range of applications. In this context, we use atomic-scale (Molecular Dynamics) and continuous (Finite Elements) calculations to investigate in details gold nanoparticles under deformation. By combining these two approaches, we show that the elastic properties of such nano-objects are driven by their size but, above all, by their shape. This outcome was achieved by introducing a descriptor in the analysis of our results enabling to distinguish among the different nanoparticle shapes studied in the present work.
In addition, other transition-metal nanoparticles have been considered (copper and platinum) using the aforementioned approach. The same strong dependence of the elastic properties with the shape was revealed, thus highlighting the universal character of our achievements. 

\end{abstract}


\maketitle


\section{Introduction}
\label{sec:Introduction}

Metallic nanoparticles (NPs) are fascinating objects with unique properties that differ significantly from their bulk counterparts due to their high surface area to volume ratio. Consequently, many nanoparticle-based technology approaches are foreseen in different fields such as catalysis, medical or optics applications~\cite{He2018}. However, NPs can be subjected to mechanical constraints, whatever their domain of use, leading to structural modifications or even irreversible changes that can drastically  affect their intended application~\cite{Guo2013}. In this context, understanding the structural modifications of small-sized metallic NPs induced by elastic deformations is crucial. Yet, this must be adressed by taking into account that the physics at the nanoscale is particular, especially for NPs whose diameter is smaller than around 10 nm. Typical examples include the dependence on the size of the melting temperature~\cite{Buffat1976} or surface energy~\cite{Amara2022} in case of pure NPs as well as the order-disorder transition temperature for bimetallic nanoalloys~\cite{Alloyeau2009}. 

Regarding the mechanical properties of metallic materials, an increase in strength is commonly observed when the characteristic microstructural length scale is reduced~\cite{Uchic2004, Wu2005}. This mainly concerns systems with a minimal size of the order of a hundred nanometers where three regimes have been highlighted In case of plastic deformations: the nanometer regime ($\sim$ 100 nm and below), an intermediate regime (between 100 nm and approximately 1 $\mu m$), and a bulk-like regime~\cite{Kraft2010}. However, the study of the mechanical properties of NPs of a few nanometers is much more complex. From an experimental point of view, in situ Transmission Electron Microscopy nanoindentation is a suitable tool for addressing this issue as it provides real-time monitoring of the deformation~\cite{Deneen2006, Carlton2012, Legros2014, Abad2021}. Although powerful, these techniques are rather cumbersome limiting their use for systematic studies and often require a numerical support. Indeed, different theoretical models based on analytical calculations have been proposed to characterize the structural properties of NPs under elastic deformation~\cite{Cuenot2004,Shaat2019}. Despite being attractive on a purely formal point of view, these models are not really adapted to NPs since they do not take explicitly into account their finite shape i.e. the prevalent role of the surfaces, the vertices and the edges that lead to specific structural properties. More specifically, the nature of the different surface sites should not be ignored, especially since as the size decreases, the contribution of facets, which are predominant for large dimensions, decreases in favour of vertices and edges. In order to obtain an atomic-scale view, in which the surface sites are clearly treated, computer simulations are generally considered very useful tools to investigate the mechanical properties of NPs. Among the possible techniques~\cite{Cherian2010,Pizzagalli2020,Amodeo2021}, molecular dynamics (MD) simulations are the most widespread as they enable to precisely characterize the elastic and plastic mechanisms of NPs under deformation whose sizes are close to the experiments~\cite{Mordehai2011,Feruz2016,Kilymis2018,Abad2023}. 

Surprisingly, there is no exhaustive study regarding the effects of size and shape on the elastic properties of NPs. Most of the previously published works concerned uniaxial compression along a specific orientation for fairly large NPs (above 10 nm) whose shape is rarely observed (cubic, spherical, truncated spherical, ...). However, these studies are incomplete to better understand and predict the properties of metallic NPs under elastic deformation. To address this issue, different sizes (ranging from 4 to 25 nm) and shapes (truncated-cube, cubo-octahedra and Wulff) of NPs, close to those experimentally observed, are investigated using complementary approaches, i.e. MD and Finite elements (FE). Thanks to these approaches, the elastic deformations during nanoindentation along the (111) and (001) facets are analyzed. NPs of three different metal (Au,Cu,Pt) will be considered to draw conclusions as general as possible.

\section{Methodology}
\label{sec:Methods}
\subsection{Nanoparticles}

\begin{figure}[htbp!]
\includegraphics[width=0.8\linewidth]{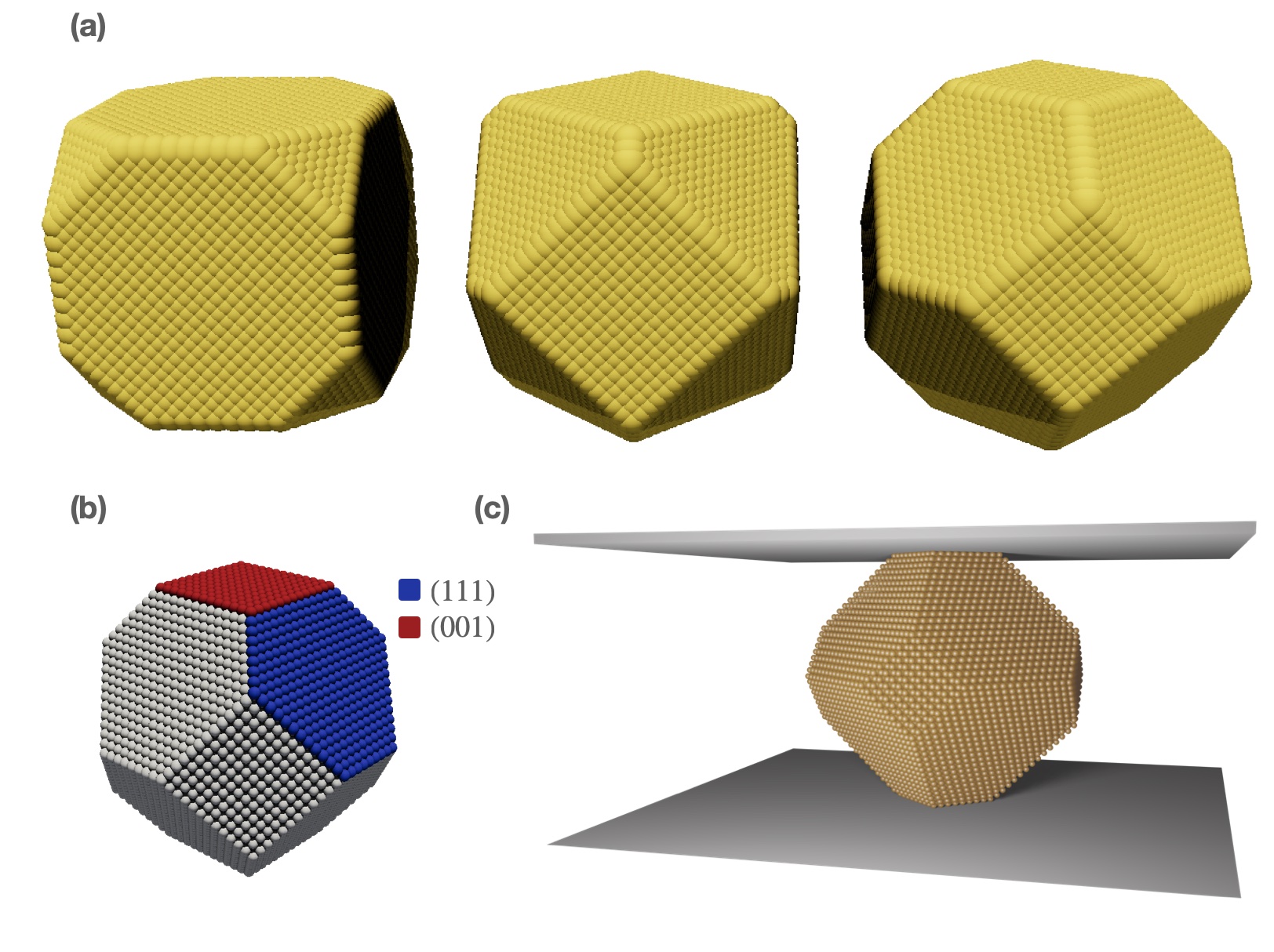}
\caption{(a) Truncated cube, cuboctahedral and Wulff structures  (b) (001) and (111) facets present at the surface of the NPs. (c) Schematic representation of the indentation of the NP where plane indenters are imposed.}
\label{fig:fig_1}
\end{figure}

Nanoparticles are systems of finite size that cannot be considered as simple fragments of a crystalline solid thus adopting structural arrangements that differ from the those of bulk~\cite{Henry2005}. In this context, Wulff established a well known rule to describe the equilibrium forms of  free polyhedral crystals~\cite{wulff1901} that has been extended to supported systems~\cite{Kaischew1952}. Among all the possible identified structures, we focus our study on morphologies that are observed experimentally , i.e. truncated cube, cuboctahedral and Wulff shapes (see Fig.~\ref{fig:fig_1}a). Note that the particular case of icosahedral NPs, common for small aggregates, is not considered here since the icosahedral arrangement is unstable for particles larger than 4-5 nanometers.
As seen in Fig.~\ref{fig:fig_1}b, the shapes addressed in the present work are based on the face-centered cubic (FCC) lattice, leading to the presence of outer facets of (111) and (001) orientations. To highlight size effects, NPs with diameters ($d$) ranging from 4 to 25 nm range (around $10^2$  to $10^6$ atoms, respectively) are taken into account. More precisely, $d$ is defined as following: 
\begin{equation}
    d = 2 \cdot \left({\frac{3V_{eff}}{4\pi}}\right)^{1/3} \nonumber
\end{equation}
with $V_{eff}$ being the effective volume of a sphere equal to the exact volume of the considered nanoparticle. The size is strictly related to the number of atoms of the NP and allows us a more methodical study of different shapes at same size. Lastly, different transition metal elements (Au, Cu and Pt) are considered to generalize as much as possible our conclusions on the elastic properties. \\

In the following, the NPs deformation procedure using plane indenters (see Fig.~\ref{fig:fig_1}c) is detailed from atomistic and continuous approaches.

\subsection{Atomistic calculations - Molecular Dynamics}

The MD simulations are carried out using the open source Large-scale Atomic/Molecular Massively Parallel Simulators (LAMMPS) package~\cite{Plimpton1995}. To study nano-indentation of NPs,  we apply the methodology successfully adopted in Ref.~\cite{Mordehai2011, Kilymis2018, Roy2019, Abad2023}. In this study, we explore Au, Cu and Pt NPs at the atomic level by using a specific $N$-body inter-atomic potential derived from the second moment approximation of the tight-binding scheme (TB-SMA).~\cite{Ducastelle1970, Rosato1989} 
\begin{table}[!ht]
\centering
\begin{tabular}{c|c|c|c|c|c}
\hline
\hline
                      & Lattice       & Cohesive   &B	&C'    &C$_{44}$  \\
                      & parameter & energy 	     &         &        &                 \\ \hline
SMA (Au)              & 4.08          & -3.81 	     & 165  &16.0    &44.5             \\ 
 Experiments (Au) & 4.08          & -3.81	     & 166  & 16.0   & 45.0            \\ \hline
SMA (Cu)              & 3.63         & -3.39 	     & 154  &23.0    &54.5             \\ 
 Experiments (Cu) & 3.62          & -3.50	     & 142  & 24.0   & 75.0           \\ \hline
SMA (Pt)              & 3.98          & -5.53 	     & 220  &35.0    &100             \\ 
 Experiments (Pt) & 3.92          & -5.86	     & 288  & 52.0   & 77.0            \\ 
\hline
 \hline
\end{tabular}
\caption{Comparison of our TB-SMA model with experimental data: lattice parameters (\AA)~\cite{Kittel1995} cohesive energies (eV/at),~\cite{Kittel1995} bulk modulus and elastic modulii (GPa)~\cite{Simmons1971} for Au, Cu and Pt in a FCC structure.}
\label{tab:SMA}
\end{table}
The accuracy of TB-SMA models depends critically on the validity of the database to which the parameters of the potentials are fitted. Here, we summarize only the crucial points of the TB-SMA potential and refer the reader to the Ref.~\cite{Delfour2009, Chmielewski2018, Front2021} for more details. In particular, the parameters of the TB-SMA potentials are obtained by fitting on the experimental values of the lattice parameter, the cohesive energy, and the elastic moduli (bulk modulus and the two shear moduli) for a FCC structure as seen in Table~\ref{tab:SMA}. Such approach is then perfectly adapted to deal with large systems ($\sim$1000 to 10000 atoms) and to reproduce the main energetic properties of transition and noble metals. Using this approach, it is interesting to note that we are able to distinguish the different surface sites, i.e. vertices, edges and facets. Obviously, with increasing size the contribution of the vertices and edges, which are preponderant for small sizes, decreases in favour of the facets.
\begin{figure}
\includegraphics[width=1.0\linewidth]{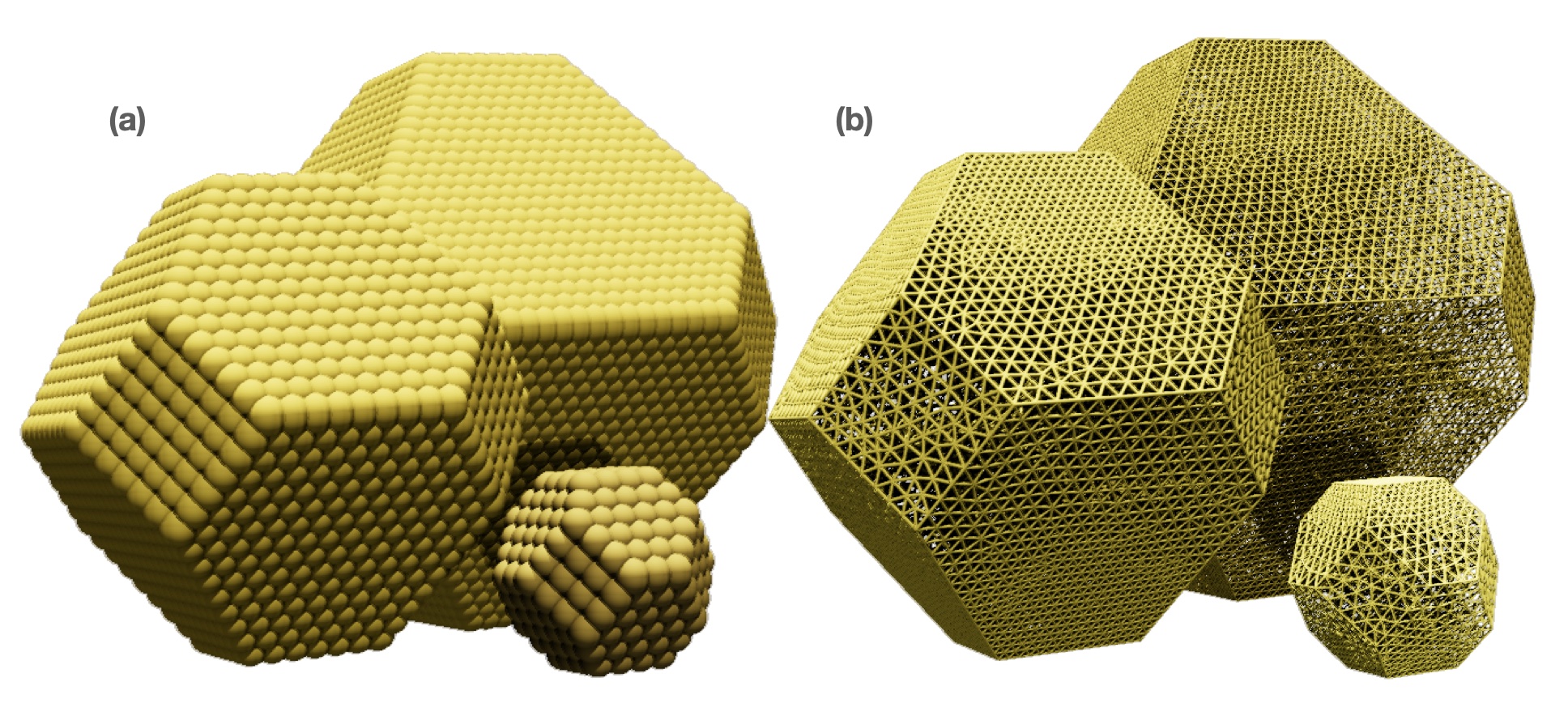}
\caption{(a) NP of different sizes revealing the surface sites i.e. vertices, edges,  (001) and (111) facets. (b) Typical meshes used in FE calculations where the mesh is more refined on the surface sites.}
\label{fig:fig_2} 
\end{figure}

To model NPs  compression via the nanoindentation procedure, the Verlet algorithm with a timestep of 1 fs combined with the Nose-Hoover thermostat at 0.01 K is used for the integration of equations of motion in the NVT canonical ensemble.  The size of the supercell in the all directions is large enough to avoid artefacts due to interaction between images caused by periodic boundary conditions. Prior to compression, the NPs are relaxed at 0.01 K until convergence of the total energy (a relative error of $\epsilon_r = 10^{-5}$ has been considered) to get the equilibrium configuration. As seen in Fig.~\ref{fig:fig_1}c, the indenter and substrate are simulated with two rigid infinite planes, parallel to the (001) or (111) facet of the NP. The effect of an indenter is implemented by introducing in the MD simulation fictitious repulsive forces to be integrated. In the present simulations, we used a quadratic repulsive force: $F(r) = -K (r-R)^2$ with $K = 1000$ eV.\AA$^{-3}$ and $R$ corresponding to the indenter position in agreement with previous works~\cite{Mordehai2011, Roy2019}. During the simulation we use two moving indenters (bottom and top plane in Fig.~\ref{fig:fig_1}c), each of them applying a strain rate of about $3\cdot 10^{7}$s$^{-1}$ which corresponds to an indenter velocity of roughly  0.06 to 0.6 m.s$^{-1}$  As stated in the Ref.~\cite{Mordehai2011}, the velocity has to be less than the speed of sound in the considered material to allow the atoms to reorganise before a new displacement is imposed.

\subsection{Continuous modeling - Finite Element}
In our continuous approach, indentation has been modeled using FE method assuming a linear elastic material. Within this framework, the stress acting on a system is directly proportional to its deformation and can be expressed by the Hooke's law so that $\sigma = E \cdot \epsilon$, where $\sigma$ and $\epsilon$ are, respectively, the second order stress and strain tensor and $E$ is the elastic constant third order tensor. In FE calculations, the linear elastic problem can be solved by computing the mechanical equilibrium, using the Hooke's law, in a discrete domain (here the NP), imposing the proper boundary conditions. In this study, we used an open source FE code named FEniCS~\cite{Fenics} where second order partial differential equations and a variational approach are adapted to handle a linear elastic problem. The continuous domain is discretised in a three dimensional mesh, where the solution, the displacement field, is computed at the nodes (see Fig.~\ref{fig:fig_2}a). It is worth noting that the generation of an accurate mesh is a fundamental step for obtaining results comparable with atomistic simulations. With this purpose, the meshes are refined until the results converge by means of a linear iterative solver. Similar to atomic-scale MD studies, a displacement is applied as a boundary conditions, to mimic a rigid indenter, at the top surface of the NP while the bottom surface is fixed. A non uniform mesh is generated with more points on the regions near the contact surfaces (i.e. the top and the bottom surfaces), where more heterogeneous stress field is expected, as seen in Fig.~\ref{fig:fig_2}a. Moreover, the resolution of the continuous equations implies to consider the proper elastic constants of the studied system.  In this case, the cubic symmetry of the bulk FCC crystal imposes that only three independent elastic constants are expected (c$_{11}$, c$_{12}$ and c$_{44}$). To ensure consistency with the MD simulations, the values considered here are those calculated at the atomic scale from the TB-SMA potentials (see Table~\ref{tab:SMA}), using anisotropic formulation of the Hooke's law.

\section{Gold nanoparticles}

The elastic properties of gold NPs calculated by the two previous approaches are now compared macroscopically and locally. First, the true stress true strain curves are investigated to identify the overall mechanical properties. In a second step, the stress maps are calculated to have a local insight of the elastic deformations of NPs. \\

\subsection{\label{sec:stress-strain}Stress - strain curves}

\begin{figure}[htbp!]
\includegraphics[width=1.0\linewidth]{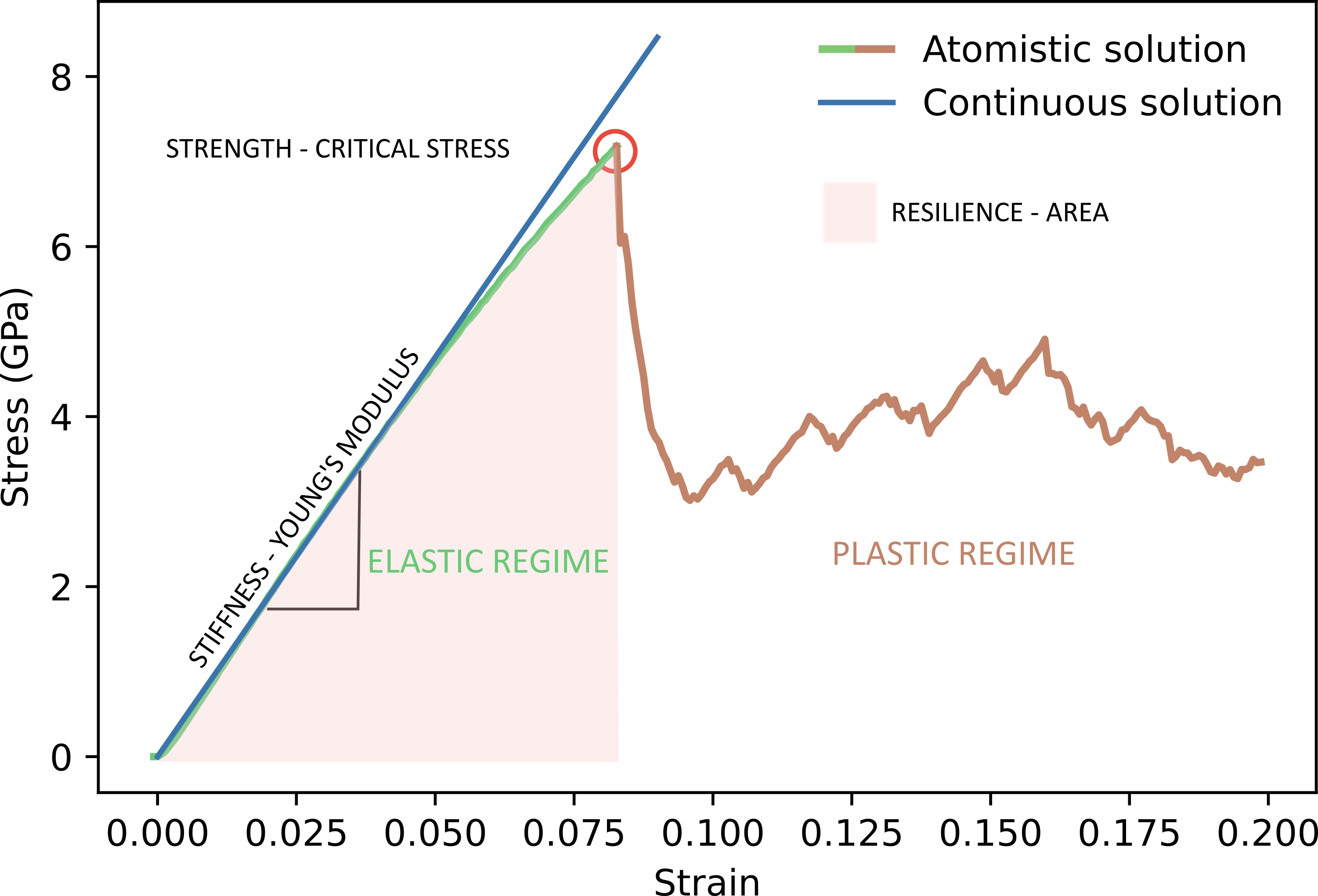}
\caption{Typical stress-strain curve. Comparison between the continuous solution and the atomistic one.}
\label{fig:fig_3new}
\end{figure}

We first examine the mechanical response of NPs under deformation by means of stress-strain curves. Figure~\ref{fig:fig_3new} shows a typical example of this analysis for a given size and shape of Au NP. 
\begin{figure}[htbp!]
\includegraphics[width=1.0\linewidth]{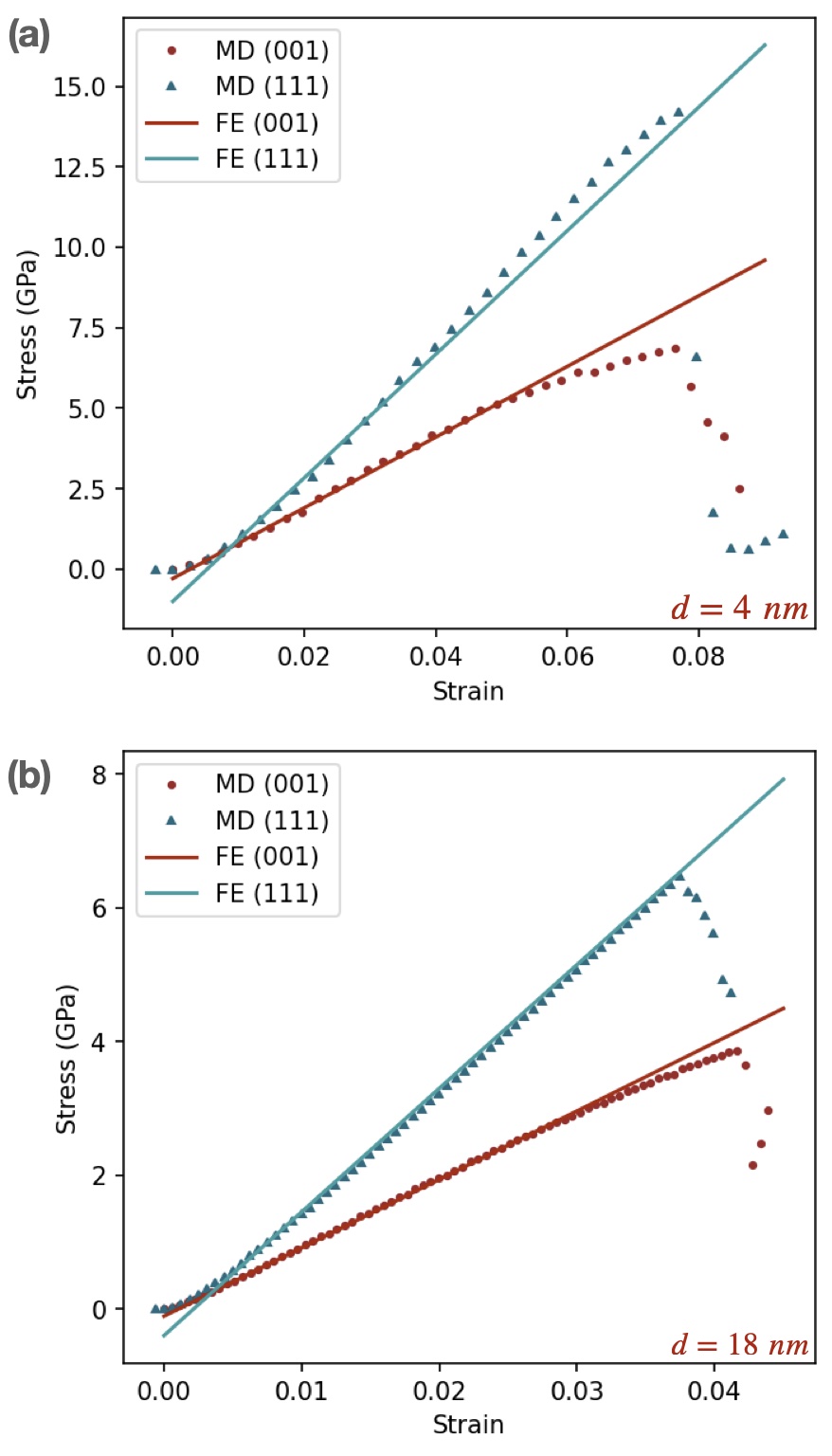}
\caption{Compressive stress as a function of strain of Wulff-shaped NP of different sizes: (a) 4 nm and (b) 18 nm. Comparison between FE and MD calculations with indentation on (001) and (111) facets.}
\label{fig:fig_4}
\end{figure} 
Regarding the MD simulations, the characteristics common to all the calculated results show two main regimes. During the very early stages of the loading, the stress variation is initially linear corresponding to an elastic regime. Then, the stress has a peak value, which defines the yield point corresponding to the largest stress that the NP can handle. Above this point, a meaningful reduction of the stress is noticed, which indicates that the plastic deformation has been activated. More precisely, the emergence of plasticity in NPs results from the heterogeneous nucleation of dislocations from the contact surfaces~\cite{Mordehai2011, Kilymis2018, Roy2019}. In the case of FE calculations, it can be noticed that only the elastic part is probed since FE formulation is based on the Hooke's law (see Fig.~\ref{fig:fig_3new}). By focusing on the initial deformation process, elastic properties can be extracted. To this purpose, it is therefore necessary to determine a macroscopic quantity to analyze the global behavior of the NP under elastic deformation. As done in previous works \cite{Mordehai2011, Kilymis2018, Roy2019}, we define an average value named the true stress compression equal to the ratio between the force applied by the indenter and the surface contact area ($A)$. At the atomic level, $A$ is not a well defined concept. As a result, many approaches exist but all of them follow the same recipe~\cite{Jacobs2017}. Firstly, we need to identify the atoms in contact with the indenter. In our case, all atoms at a distance lower than 1 \AA~from the considered indenter. It is noteworthy that no significant differences are reported by slightly varying this parameter. Then, $A$ is determined by computing a Delaunay triangulation of the points defined by the coordinates of atom centers at each indenter displacement. Consequently, the true stress strain in MD calculations can be obtained, since the indenter position is a known quantity. From FE calculations, the true strain stress is an easy accessible data. The only difficulty lays in the force calculation which is not straightforward and this has to be done by computing the reaction force generated by the NP  on the surface where the displacement is imposed.

To highlight elastic size effect, the mechanical responses of  NPs of different diameters $d$ are considered. Figure~\ref{fig:fig_4} depicts the stress-strain curves of a small ($d=4$ nm) and larger ($d=18$ nm) Wulff NPs nanoindented with plane indenter on the (001) and (111) facets. In the case of MD simulations, deviation from linearity in the elastic regime is observed for the small NP as seen in Fig.~\ref{fig:fig_4}a. This trend is found for all particles below 5-6 nm in agreement with previous calculations \cite{Feruz2016,Amodeo2017,Kilymis2018} where the influence of surfaces on the mechanical behavior inside small NPs is emphasized. For the larger NPs, a linear trend is observed in the elastic regime until plastic deformation, whatever the facet considered (see Fig.~\ref{fig:fig_4}b). Obviously, finite elements fail to reproduce the deviation from the linear regime observed for small particles (intrinsically, no deviation from linearity can be observed). In contrast, the agreement between MD and finite elements for larger particles is quite remarkable proving that a linear elasticity framework can be used to model mechanical properties at the nanoscale. Besides, plastic behaviour can be noticed in MD results:  the critical yield stress ($\sigma_C$) can be extracted since it corresponds to the maximum value before a drop, corresponding to dislocation nucleation events. As already discussed in~\cite{Mordehai2011}, increasing the NPs size implies a reduction in the critical stress according to the following equation: $\sigma_C = \alpha\cdot d^{-B}$ where $\alpha$ and $B$ are fitted values. We found that for the (001) and (111) facets, $B$ is equal to 0.461 and 0.523, respectively. The second value is in good agreement with the $0.530 \pm 0.013$ found by Feruz et \textit{al.} for Wulff structure~\cite{Feruz2016}. Note that so far only (111) facets were considered in the literature. Hence, we show here that this size-dependent strength can also be extended to (001) facets. A more detailed analysis on the onset of plasticity in FCC metal NPs will be presented in a future work.

\begin{figure}[htbp!]
\includegraphics[width=1.0\linewidth]{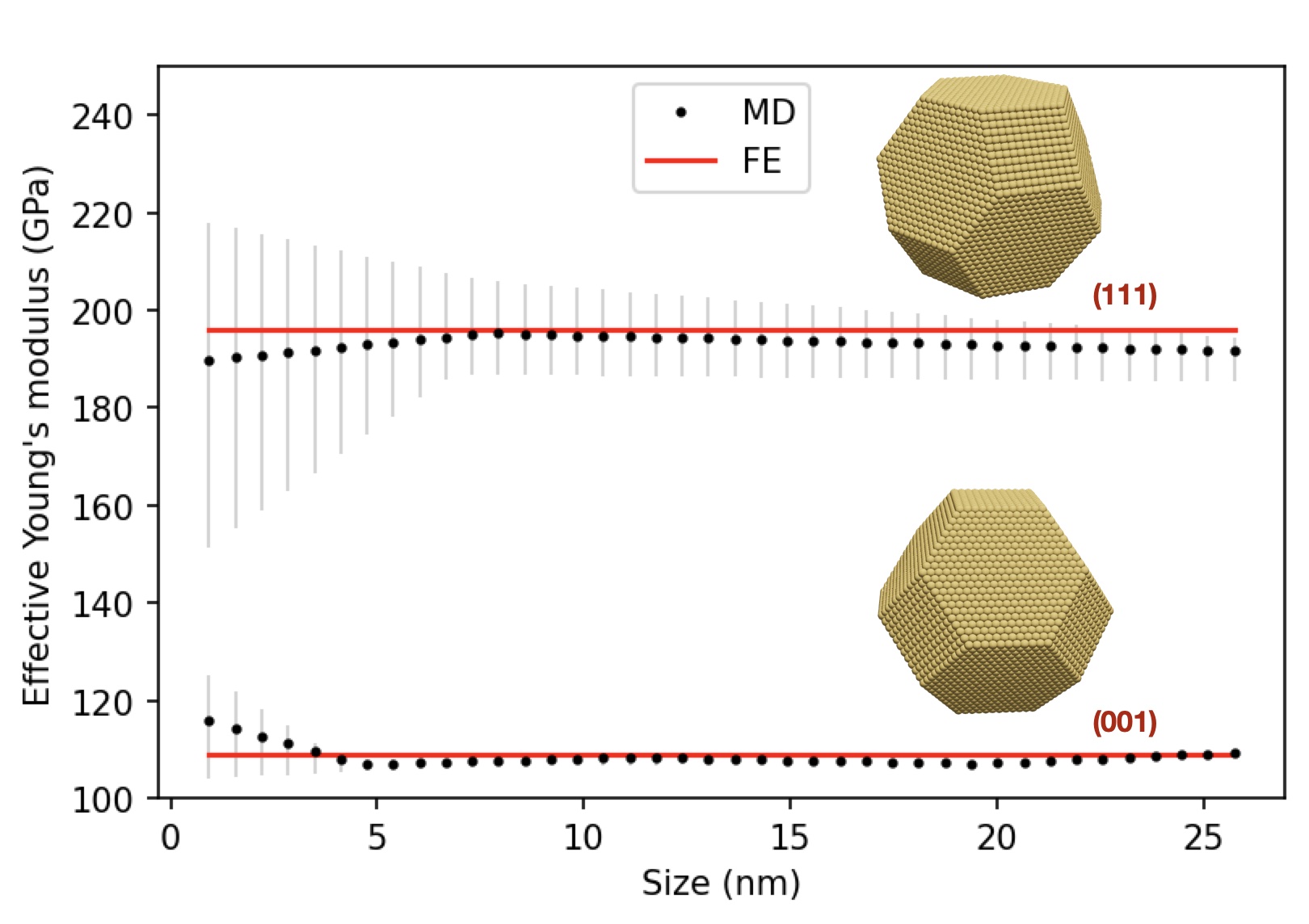}
\caption{Effective Young modulus as a function of the size for Wulff-shaped gold nanoparticles. Comparison between continuous and atomistic calculations with indentation on (001) and (111) facets.}
\label{fig:fig_5}
\end{figure}

\begin{figure*}[htbp!]
\includegraphics[width=0.85\linewidth]{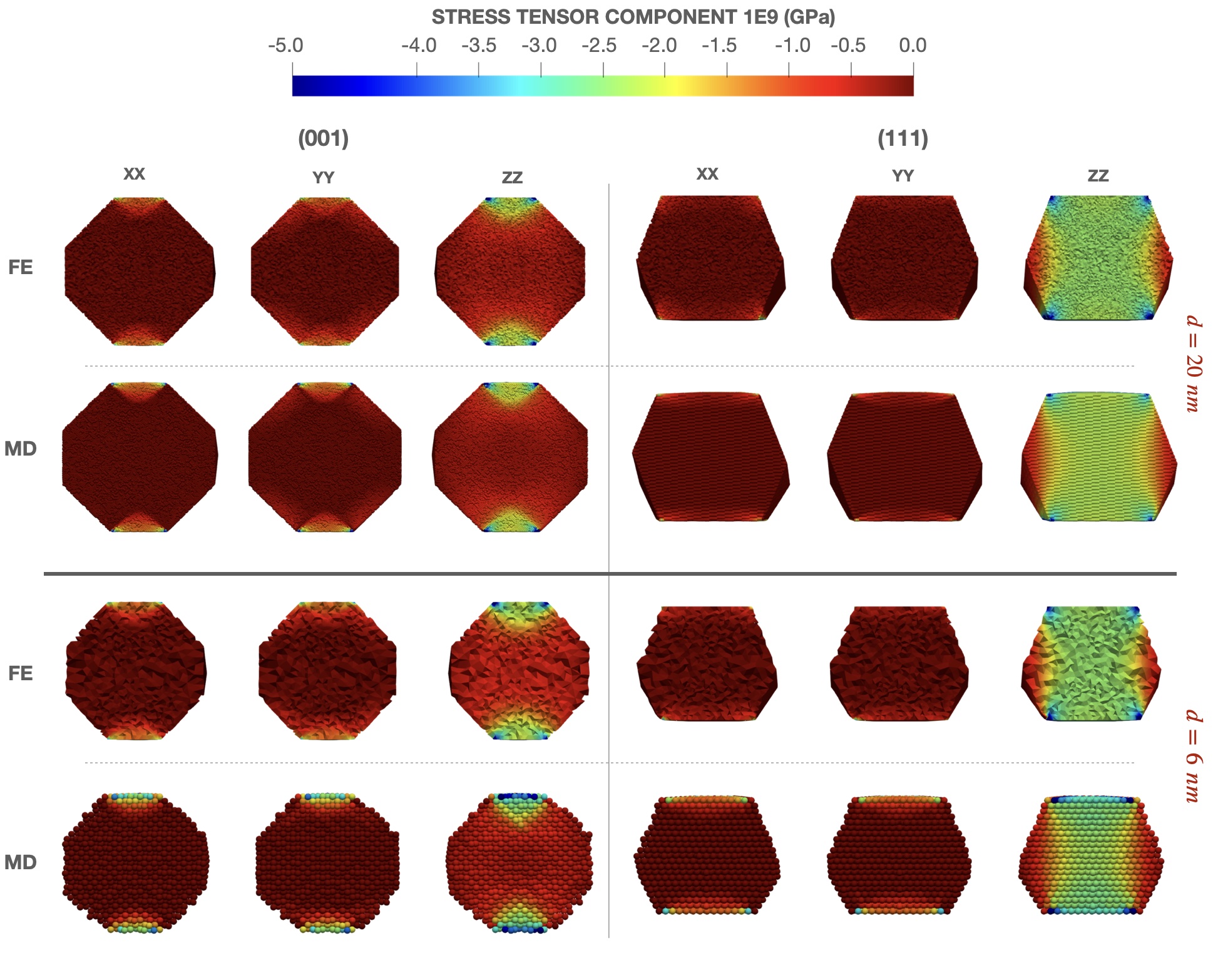}
\caption{\label{fig:stressstraincomparison} Stress maps calculated with MD and FE approaches for Wulff-shaped gold nanoparticles of (top) 20 nm and (bottom) 6 nm diameter NPs and by imposing nanoindentation on (001) and (111) facets.}
\end{figure*}

To better analyse the elastic behaviour of NPs, effective Young modulus ($E_{eff}$ i.e. $\Delta\sigma /\Delta \epsilon$ ) are extracted from the stress-strain curves for all NP sizes. The results for indentation on both facets are presented in Fig. \ref{fig:fig_5}. Above $5$ nm, it can be observed that MD and FE calculations provide very similar results with extremely low errors. This is not so surprising for large particles since finite elements reproduce the elastic domain very well as revealed in Fig.~\ref{fig:fig_4}b. For smaller particles, it is clear that the deviation from linearity already discussed in Fig.~\ref{fig:fig_4}b is not reproduced with FE calculations. Furthermore, the parameter $E_{eff}$ is also difficult to be defined for small Nps. For this reason an error bar is introduced that consider different possible linear fit for the elastic part of a given nanoparticle stress-strain curve. Due to the deviation from linearity using MD calculations, $E_{eff}$ values for small NP have larger error bars than the large ones. Nevertheless, when comparing results from FE and MD calculations even for small NPs, the global slope of the curve is rather correct enabling to get effective Young modulus values in agreement between both approaches.  As the size of the nanoparticles increase the value of the effective Young modulus for both (001) and (111) indentation converges to a plateau \cite{Armstrong2012}. while the $E_{eff}^{(111)}$ is larger than $E_{eff}^{(001)}$, as expected in anisotropic media ($E^{(111)} > E^{(001)}$ for bulk),  the value of the plateau depends on the shape of the NP, as it will be highlighted in section \ref{sec:shape_effect}.


\subsection{Stress Map}

In the previous section we discussed the mechanical properties of NPs from a global point view   via the stress-strain curve. Here we want to analyse local elastic properties, by the means of stress maps, comparing MD and FE outcomes. In order to highlight the similarities between the two simulation approaches, the surface contraction peculiar to transition metals~\cite{Gupta1981} reproduced in atomistic simulations is neglected. Therefore, we took as reference state (no stress in the NP) the atomic positions after relaxation steps. Concerning the FE approach, the displacement fields is computed at each node of the mesh for an imposed indentation depth. From this displacement field, the stress is computed and by interpolation is projected on the mesh nodes. Obviously, the mesh size is crucial to achieve robust results. Therefore, the mesh was refined until the convergence of stress values. In Fig.~\ref{fig:stressstraincomparison}, the stress distributions obtained from MD and FE calculations are presented in case of small (6 nm) and large (20 nm) NPs. This corresponds to nanoindentations on a (001) facet (on the left) and on a (111) facet (on the right) at $2.5\%$ strain, in a linear regime. Concerning the (001) indentation, stress is concentrated in two conical regions, close to the plane indenters. For all the components of the stress field, the maximum value is observed at the corners as already discussed in \cite{Roy2019,Mordehai2011}. If we observe the ZZ component, the stress field is negative due to the compression imposed by the applied loading and we can notice that with the presence of free surfaces, the stress decrease rapidly moving from the top or the bottom surface to the center of the nanoparticle. Remarkably, FE and MD calculations give very similar results for both NP sizes: the values for stress tensor components computed with the two approaches are very close, with differences less than 5 \%. Such common features were also obtained for a nano-indentation on the (111) facets, for any particle sizes and shapes. The only dissimilarity with (001) indentation is that the stress distribution is more homogeneous in the z direction. This is because the plane indenters has a larger contact area, indeed for Wulff structures (111) surfaces are wider then (001). \\

Studying nanoindentation of a Wulff Au nanoparticle, we have shown both finite elements and atomistic simulations are perfectly adapted to capture the elastic properties on both a local (stess map) and a global (stress-strain curve) viewpoint. The agreement betwween the two different approaches is good, notably for NPs with diameter above 5 nm. It is then tempting to think that such features are mainly driven by the shape of the particle and not by atomistic features such as the presence of specific surface sites (edges, vertices and surface arrangement).

\section{\label{sec:shape_effect} Shape effects}

\begin{figure}[htbp!]
\includegraphics[width=1.0\linewidth]{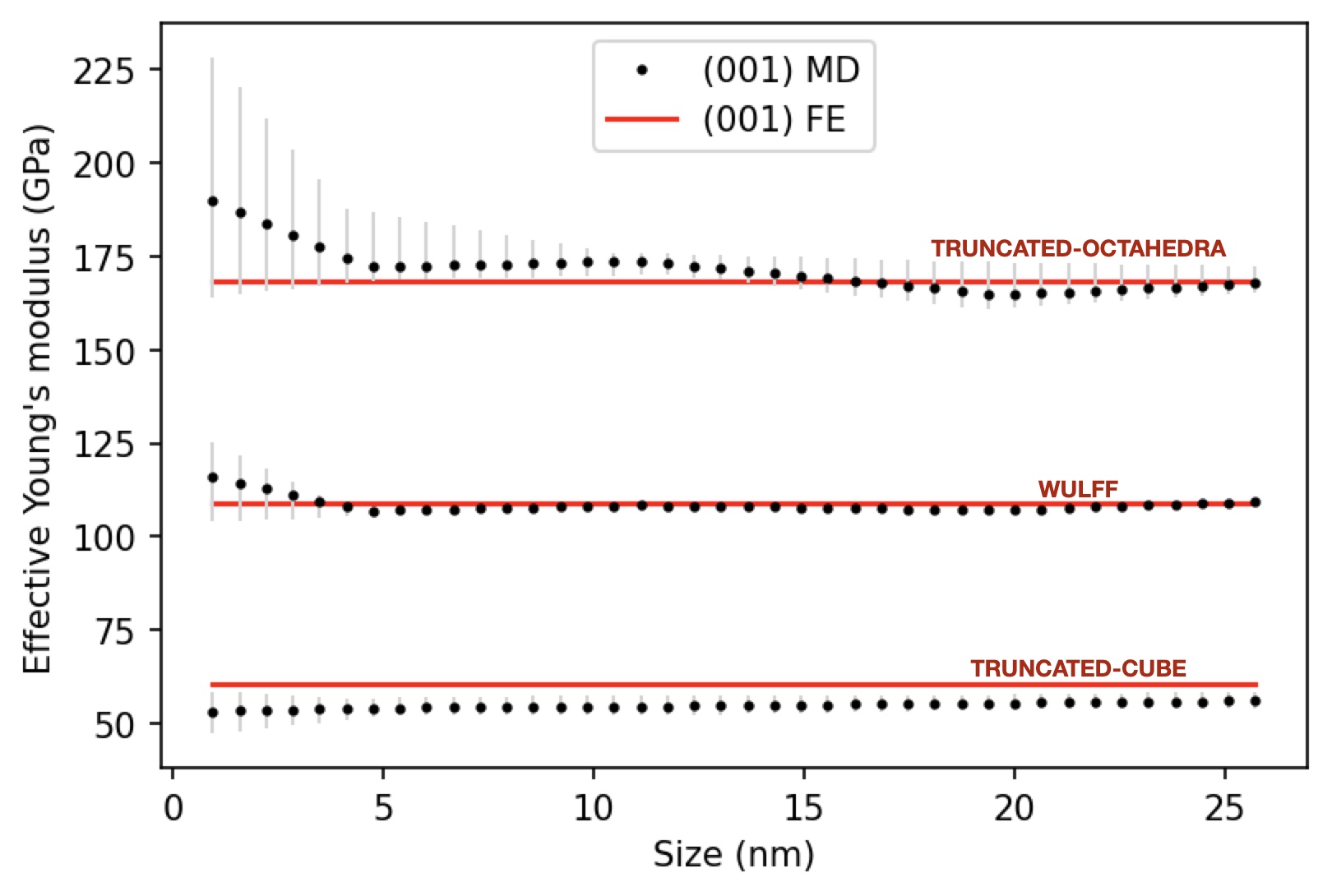}
\caption{Effective Young modulus as a function of the size for gold nanoparticles with different shapes. Comparison between continuous and atomistic calculations with indentation on (001) facets.}
\label{fig:fig_7bis}
\end{figure}

To  highlight  shape  effect on the elastic properties,  the  mechanical  responses  of different Au NPs  are considered. In Fig.~\ref{fig:fig_7bis}, the calculated effective Young modulus for various NP shape as a function of size are presented. 
First, we can notice that the general observation of section~\ref{sec:stress-strain} for Wulff shape NPs are valid also for truncated-octahedra and truncated cubes: for NPs larger than 5 nm the value of $E_{eff}$ converge to a plateau and no visible size effect is observed. In the case of the truncated-octahedron, there is a stronger variation of the $E_{eff}$ values for small size NPs ($<5$ nm) in MD calculations, from $\sim$ 200 to $\sim$ 175 GPa. For larger particles, $E_{eff}$ reaches a constant value around 175 GPa. Again, FE approach is not able to reproduce such a size dependence, suggesting that the role of specific surface sites cannot be neglected in small NPs.
On the other hand, the agreement between FE and MD for larger particles is quite remarkable. More interestingly, a strong shape effect is observed in Fig.~\ref{fig:fig_7bis}, where the $E_{eff}$ plateau value changes depending on the shape. Indeed, wide differences are obtained ranging from $\sim$ 50 GPa (truncated cube) to $\sim$ 175 GPa (truncated octahedra). To further emphasise the observed shape effect, we introduce a descriptor $G$ to classify the shape of the NPs with respect to the ratio between (001) and (111) facets:
\begin{equation}
G=\frac{8 \cdot A_{(111)}}{6\cdot A_{(001)}},     
\end{equation} 
where $A_{(111})$ and $A_{(001)}$ are the area of the (111) and (001) facets, respectively. The factors 8 and 6 correspond to the number of free surfaces. 
\begin{figure}[htbp!]
\includegraphics[width=1.0\linewidth]{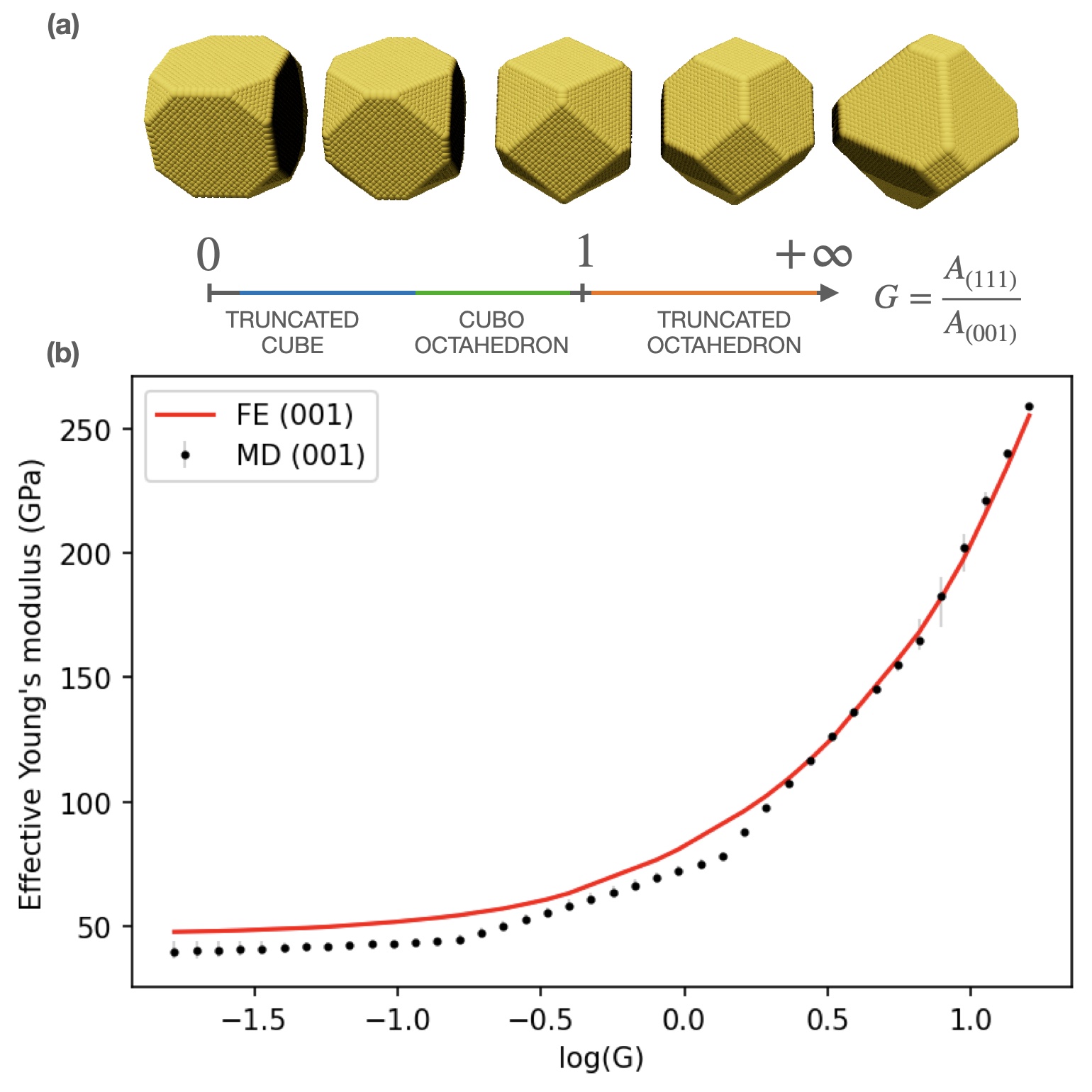}
\caption{(a) The descriptor $G$ introduced to discriminate the different shapes of NP. (b) Effective Young modulus as a function of the shape for gold nanoparticles. Results correspond to an indentation on (001) facets for a 20 nm diameter NP.}
\label{fig:fig_7tris}
\end{figure}
From this descriptor, it is then possible to clearly distinguish the different shapes considered in our study (see Fig.~\ref{fig:fig_7tris}). Moreover, two limiting cases are also identified: the cube and the octahedron with $G=0$ and $G=\infty$, respectively. At a given NP size (here 20 nm), Fig.~\ref{fig:fig_7tris}b depicts the variation of the effective Young modulus for different shapes. Again, a very good agreement is found between FE and MD calculations. A maximum difference of about 4$\%$ is revealed and corresponds to  the truncated-cube shapes. Such deviation is due to the non linearity of the stress-strain curves in approaching a cubic system (for cubes a very large elastic deformation of about $10 \%$  is observed before the plastic onset, so non-linear behaviour is observed in MD simulations).  
\begin{figure*}[htbp!]
\includegraphics[width=0.85\linewidth]{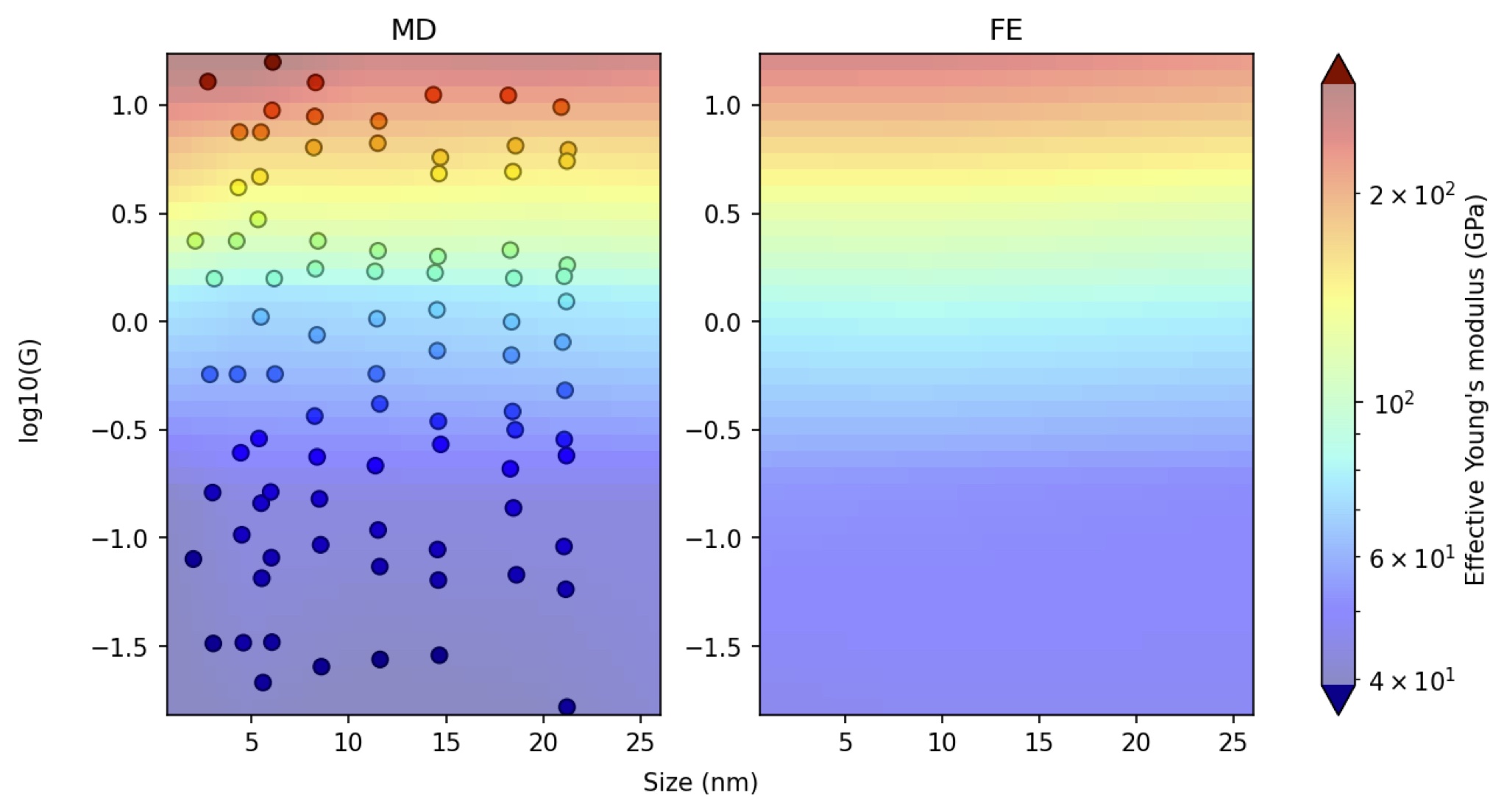}
\caption{Effective Young modulus as a function of the size and shapes for gold nanoparticles plotted on a bi-dimensional map. Comparison between atomistic (on the left) and continuous (on the right) calculations for an indentation on (001) facets.}
\label{fig:fig_map}
\end{figure*}
Very interestingly, a significant variation of the order of 200 GPa can be observed between the two limiting shapes. This considerable effect illustrates that the influence of the NP shapes on the elastic properties is huge for a 20 nm nanoparticle. \\

To generalize, $E_{eff}$ computed values in function of  different shapes ($G$ descriptor) and of NP sizes ranging from 4 to 25 nm is presented in Fig.~\ref{fig:fig_map} in a bi-dimensional map for MD simulations (left panel) for FE simulation (right panel). On one hand, no size effect is identified (above 5nm). Indeed, the variation of the $E_{eff}$ with the size of the NP is negligible, never exceeding few GPa for a given NP shape. On the other hand, significant variation of the effective young's modulus are remarked with NP shape. From truncated-cube to truncated-octahedron structures, a difference of about 150 GPa on the effective young's modulus is reported. Here again, we can note that the agreement between the atomistic and finite element calculations is remarkable whatever the shape and size of the NP.\\

All the results presented so far concern nano-indentation on (001) facets. Meanwhile, the same conclusions have been achieved by investigating the deformation mechanisms on (111) facets (see Sec. I. of the Supplemental Materials). It is important to stress that our analysis show that the shape, not the size, of the gold NPs has a wide impact on its macroscopic measured elastic properties. 

\section{\label{sec:Results_Co_Pt}Generalization to other systems}

\begin{figure}[htbp!]
\includegraphics[width=0.9\linewidth]{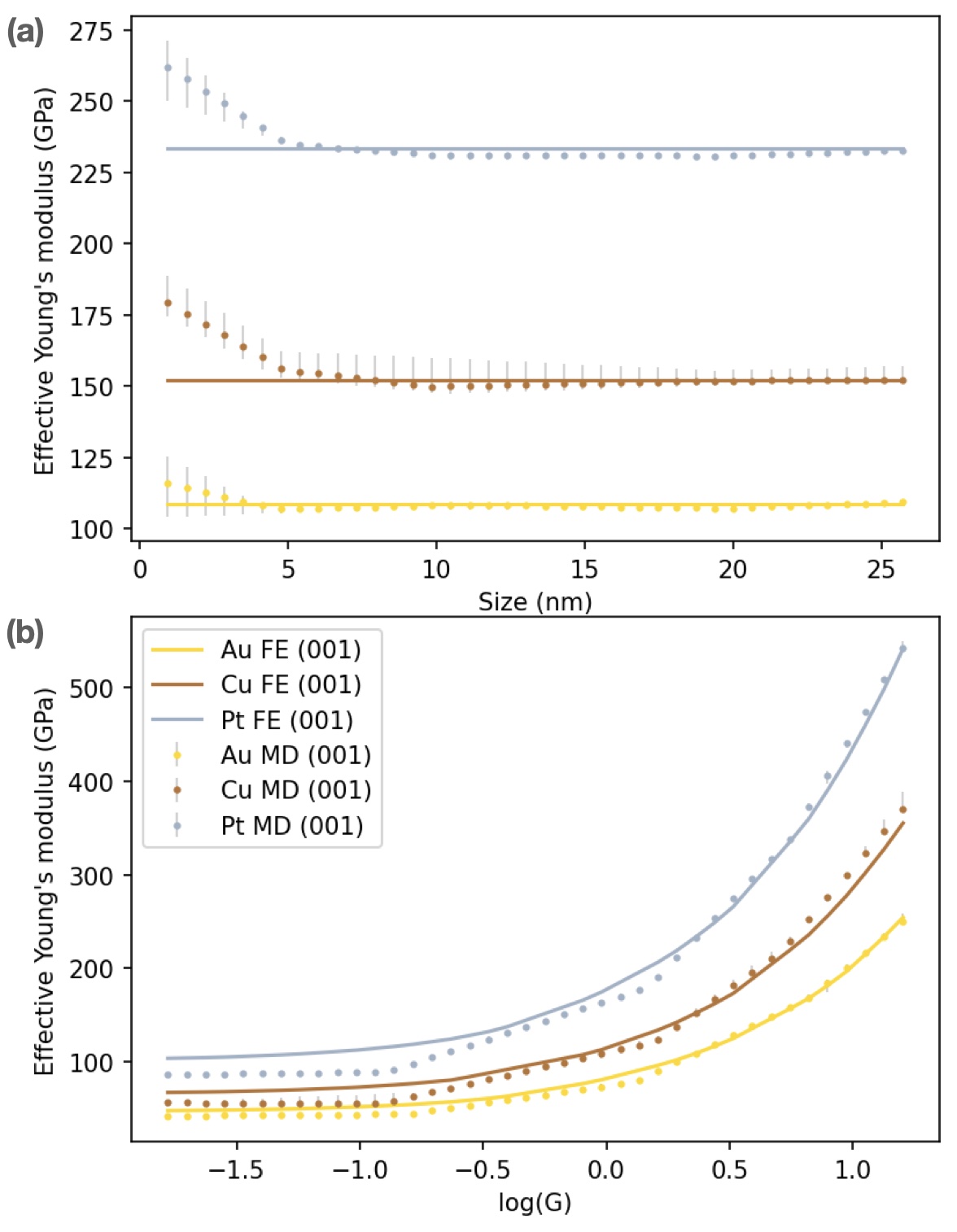}
\caption{(a) Effective Young modulus as a function of the size for Au, Cu and Pt Wulff nanoparticles. (b) Effective Young modulus as a function of the shape for Au, Cu and Pt nanoparticles of 20 nm. Comparison between continuous and atomistic calculations with indentation on (001) facets.}
\label{fig:fig_generalzation}
\end{figure}

In the previous sections, the investigations were focused on gold NPs. To generalize our conclusions, the analysis of the elastic properties of NPs has been extended to other metals, such as Cu and Pt. The calculated effective Young modulus are presented in Fig.~\ref{fig:fig_generalzation} for different sizes and shapes in case of Au, Co and Pt by imposing nanoindentation on (001) facets. More results can be found in Sec. II and III of the Supplemental Materials for Cu and Pt, respectively. 
Whatever the material considered, we first see that the finite elements reproduce the atomistic calculations for NP sizes exceeding 5 nm. For smaller diameters, the differences are fairly notable for Cu and Pt ($<40$ GPa ) but minor for Au NPs ($<15$ GPa). Beyond this limit, no size effect is reported with a constant value for $E_{eff}$. More interestingly, it can be seen that the elastic properties vary strongly with the shape of the particle. As shown in Fig.~\ref{fig:fig_generalzation}b, significant variations of the elastic properties are highlighted whatever the transition metals. Indeed, $E_{eff}$ varies of hundreds of GPa moving from small value to large value of G. For the nanoindentation of the (111) facet, the same trend is observed, and the same conclusions can be drawn for Cu and Pt, as seen in Sec. IV and V of the Supplemental Materials, respectively.

\section{Conclusion}

In this work, we have presented an extensive study on elastic deformations of metallic nanoparticles with different shapes and sizes, imposing a mechanical loading via nano-indentation on (001) and (111) facets. By combining atomistic and continuous calculations, no size effect is revealed for nanoparticles with a diameter of larger than 5 nm. More interestingly, we have shown that the elastic properties of NPs are highly driven by the shape of the particle, finding that the ratio between  (111) and (001) is a suitable descriptor to address this shape dependence. These results were achieved by nanoindenting Au, Cu and Pt NPs, showing that our conclusion can be applied to different transition metal nanoparticles. Thanks to this study, we can state that controlling the shape of nanoparticles can be viable path for the engineering of a new class of nano-objects with unique and targeted mechanical properties. \\

\section*{Supplemental material : \\Tuning Elastic Properties of Metallic Nanoparticles by Shape Controlling: \\
From Atomistic to Continuous Models}

\textbf{Sec. I. Elastic properties of Au nanoparticles under (111) deformation}\\

\begin{figure}[htbp!]
\includegraphics[width=0.85\linewidth]{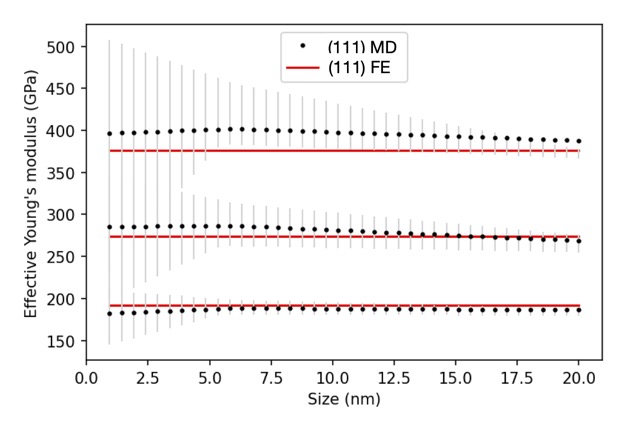}
\caption{Effective Young modulus as a function of the size for Au nanoparticles with different shapes. Comparison between continuous and atomistic calculations with indentation on (111) facets.}
\label{fig:Gold_111_1}
\end{figure}

\begin{figure}[htbp!]
\includegraphics[width=0.85\linewidth]{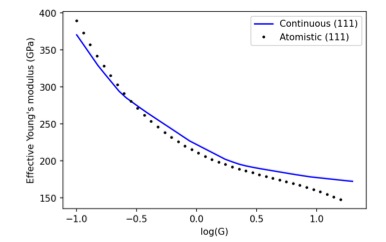}
\caption{Effective Young modulus as a function of the shape for gold nanoparticles. Results from continuous and atomistic calculations correspond to an indentation on (111) facets for a 20 nm diameter NP.}
\label{fig:Gold_111_2}
\end{figure}

\begin{figure}[htbp!]
\includegraphics[width=1.0\linewidth]{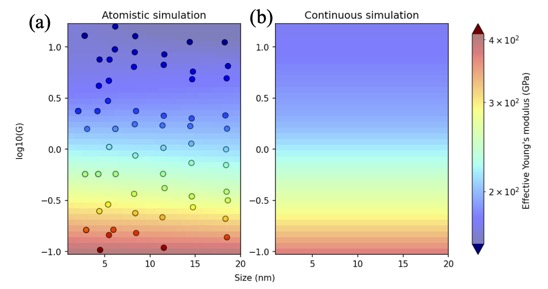}
\caption{Effective Young modulus as a function of the size and shapes for gold nanoparticles in a map form. Comparison between (a) atomistic and (b) continuous calculations with indentation on (111) facets.}
\label{fig:Gold_111_3}
\end{figure}

All the results concerning the elastic properties of gold nanoparticles for nanoindentation on (111) facets are presented in the Fig.~\ref{fig:Gold_111_1}, \ref{fig:Gold_111_2} and \ref{fig:Gold_111_3}. 
\newpage

\textbf{Sec. II. Elastic properties of Cu nanoparticles under (001) deformation}\\

\begin{figure}[htbp!]
\includegraphics[width=0.85\linewidth]{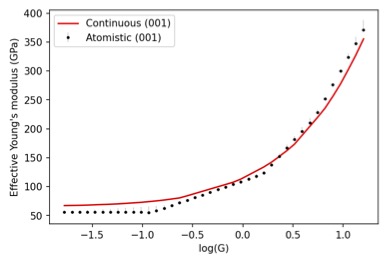}
\caption{Effective Young modulus as a function of the shape for copper nanoparticles. Results from continuous and atomistic calculations correspond to an indentation on (001) facets for a 20 nm diameter NP.}
\label{fig:Cu_001_1}
\end{figure}

\begin{figure}[htbp!]
\includegraphics[width=1.0\linewidth]{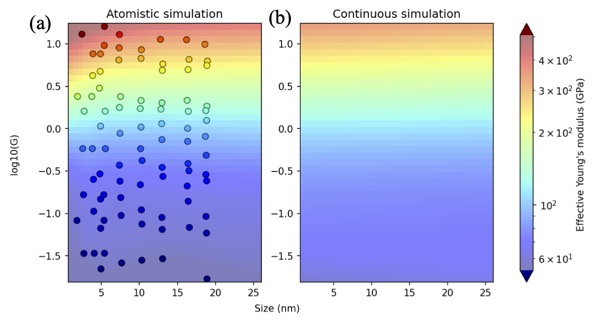}
\caption{Effective Young modulus as a function of the size and shapes for copper nanoparticles in a map form. Comparison between (a) atomistic and (b) continuous calculations with indentation on (001) facets.}
\label{fig:Cu_001_2}
\end{figure}

More results concerning the elastic properties of Copper nanoparticles for nanoindentation on (001) facets are presented in the Fig.~\ref{fig:Cu_001_1} and \ref{fig:Cu_001_2}. 
\newpage

\textbf{Sec. III. Elastic properties of Pt nanoparticles under (001) deformation} \\

\begin{figure}[htbp!]
\includegraphics[width=0.85\linewidth]{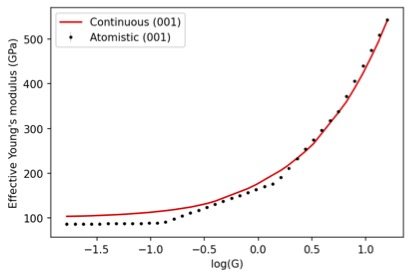}
\caption{Effective Young modulus as a function of the shape for platinum nanoparticles. Results from continuous and atomistic calculations correspond to an indentation on (001) facets for a 20 nm diameter NP.}
\label{fig:Pt_001_1}
\end{figure}

\begin{figure}[htbp!]
\includegraphics[width=1.0\linewidth]{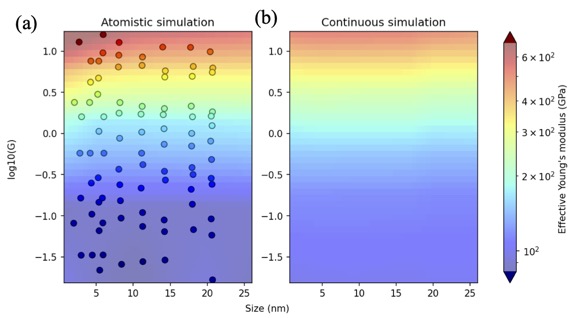}
\caption{Effective Young modulus as a function of the size and shapes for platinum nanoparticles in a map form. Comparison between (a) atomistic and (b) continuous calculations with indentation on (001) facets.}
\label{fig:Pt_001_2}
\end{figure}

More results concerning the elastic properties of Platinum nanoparticles for nanoindentation on (001) facets are presented in the Fig.~\ref{fig:Pt_001_1} and \ref{fig:Pt_001_2}. 

\newpage

\textbf{Sec. IV. Elastic properties of Cu nanoparticles under (111) deformation}\\

\begin{figure}[htbp!]
\includegraphics[width=0.85\linewidth]{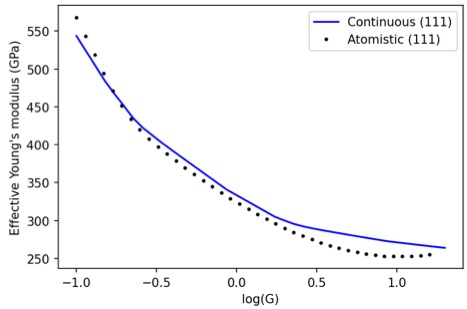}
\caption{Effective Young modulus as a function of the shape for copper nanoparticles. Results from continuous and atomistic calculations correspond to an indentation on (111) facets for a 20 nm diameter NP.}
\label{fig:Cu_111_1}
\end{figure}

\begin{figure}[htbp!]
\includegraphics[width=1.0\linewidth]{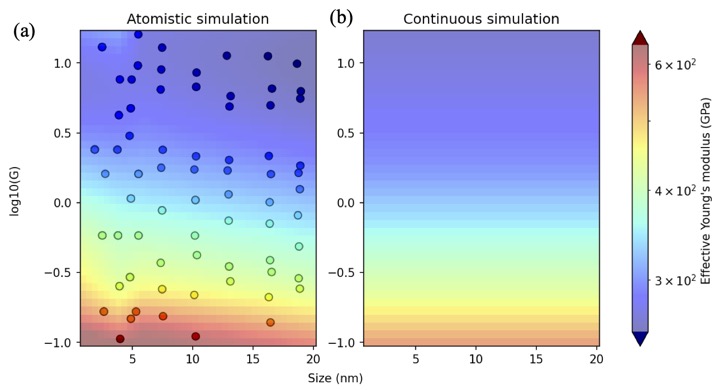}
\caption{Effective Young modulus as a function of the size and shapes for copper nanoparticles in a map form. Comparison between (a) atomistic and (b) continuous calculations with indentation on (111) facets.}
\label{fig:Cu_111_2}
\end{figure}

All the results concerning the elastic properties of Cu nanoparticles for nanoindentation on (111) facets are presented in the Fig.~\ref{fig:Cu_111_1} and \ref{fig:Cu_111_2} 
\newpage

\textbf{Sec. V. Elastic properties of Pt nanoparticles under (111) deformation}\\

\begin{figure}[htbp!]
\includegraphics[width=0.85\linewidth]{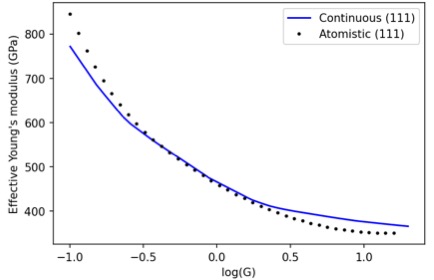}
\caption{Effective Young modulus as a function of the shape for platinum nanoparticles. Results from continuous and atomistic calculations correspond to an indentation on (111) facets for a 20 nm diameter NP.}
\label{fig:Pt_111_1}
\end{figure}

\begin{figure}[htbp!]
\includegraphics[width=1.0\linewidth]{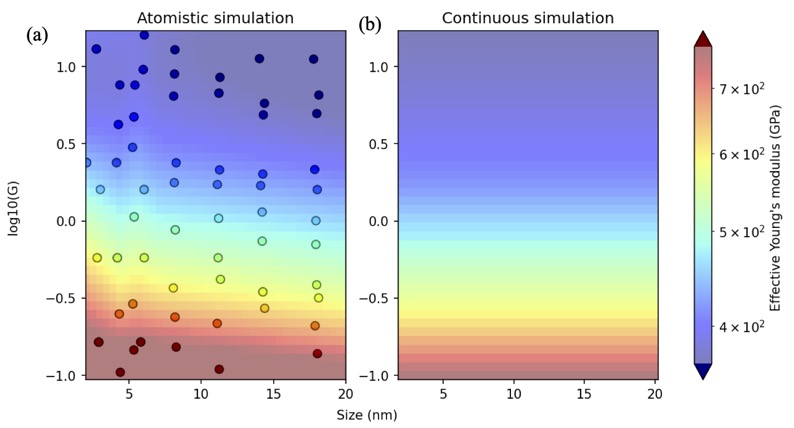}
\caption{Effective Young modulus as a function of the size and shapes for platinum nanoparticles in a map form. Comparison between (a) atomistic and (b) continuous calculations with indentation on (111) facets.}
\label{fig:Pt_111_2}
\end{figure}

All the results concerning the elastic properties of Pt nanoparticles for nanoindentation on (111) facets are presented in the Fig.~\ref{fig:Pt_111_1} and \ref{fig:Pt_111_2} 

\begin{acknowledgments}

H.A. thanks C. Ricolleau for fruitful discussions. 

\end{acknowledgments}


\begin{thebibliography}{0}%
\makeatletter
\providecommand \@ifxundefined [1]{%
 \@ifx{#1\undefined}
}%
\providecommand \@ifnum [1]{%
 \ifnum #1\expandafter \@firstoftwo
 \else \expandafter \@secondoftwo
 \fi
}%
\providecommand \@ifx [1]{%
 \ifx #1\expandafter \@firstoftwo
 \else \expandafter \@secondoftwo
 \fi
}%
\providecommand \natexlab [1]{#1}%
\providecommand \enquote  [1]{``#1''}%
\providecommand \bibnamefont  [1]{#1}%
\providecommand \bibfnamefont [1]{#1}%
\providecommand \citenamefont [1]{#1}%
\providecommand \href@noop [0]{\@secondoftwo}%
\providecommand \href [0]{\begingroup \@sanitize@url \@href}%
\providecommand \@href[1]{\@@startlink{#1}\@@href}%
\providecommand \@@href[1]{\endgroup#1\@@endlink}%
\providecommand \@sanitize@url [0]{\catcode `\\12\catcode `\$12\catcode
  `\&12\catcode `\#12\catcode `\^12\catcode `\_12\catcode `\%12\relax}%
\providecommand \@@startlink[1]{}%
\providecommand \@@endlink[0]{}%
\providecommand \url  [0]{\begingroup\@sanitize@url \@url }%
\providecommand \@url [1]{\endgroup\@href {#1}{\urlprefix }}%
\providecommand \urlprefix  [0]{URL }%
\providecommand \Eprint [0]{\href }%
\providecommand \doibase [0]{http://dx.doi.org/}%
\providecommand \selectlanguage [0]{\@gobble}%
\providecommand \bibinfo  [0]{\@secondoftwo}%
\providecommand \bibfield  [0]{\@secondoftwo}%
\providecommand \translation [1]{[#1]}%
\providecommand \BibitemOpen [0]{}%
\providecommand \bibitemStop [0]{}%
\providecommand \bibitemNoStop [0]{.\EOS\space}%
\providecommand \EOS [0]{\spacefactor3000\relax}%
\providecommand \BibitemShut  [1]{\csname bibitem#1\endcsname}%
\let\auto@bib@innerbib\@empty
\end{thebibliography}%


\begin{thebibliography}{38}%
\makeatletter
\providecommand \@ifxundefined [1]{%
 \@ifx{#1\undefined}
}%
\providecommand \@ifnum [1]{%
 \ifnum #1\expandafter \@firstoftwo
 \else \expandafter \@secondoftwo
 \fi
}%
\providecommand \@ifx [1]{%
 \ifx #1\expandafter \@firstoftwo
 \else \expandafter \@secondoftwo
 \fi
}%
\providecommand \natexlab [1]{#1}%
\providecommand \enquote  [1]{``#1''}%
\providecommand \bibnamefont  [1]{#1}%
\providecommand \bibfnamefont [1]{#1}%
\providecommand \citenamefont [1]{#1}%
\providecommand \href@noop [0]{\@secondoftwo}%
\providecommand \href [0]{\begingroup \@sanitize@url \@href}%
\providecommand \@href[1]{\@@startlink{#1}\@@href}%
\providecommand \@@href[1]{\endgroup#1\@@endlink}%
\providecommand \@sanitize@url [0]{\catcode `\\12\catcode `\$12\catcode
  `\&12\catcode `\#12\catcode `\^12\catcode `\_12\catcode `\%12\relax}%
\providecommand \@@startlink[1]{}%
\providecommand \@@endlink[0]{}%
\providecommand \url  [0]{\begingroup\@sanitize@url \@url }%
\providecommand \@url [1]{\endgroup\@href {#1}{\urlprefix }}%
\providecommand \urlprefix  [0]{URL }%
\providecommand \Eprint [0]{\href }%
\providecommand \doibase [0]{http://dx.doi.org/}%
\providecommand \selectlanguage [0]{\@gobble}%
\providecommand \bibinfo  [0]{\@secondoftwo}%
\providecommand \bibfield  [0]{\@secondoftwo}%
\providecommand \translation [1]{[#1]}%
\providecommand \BibitemOpen [0]{}%
\providecommand \bibitemStop [0]{}%
\providecommand \bibitemNoStop [0]{.\EOS\space}%
\providecommand \EOS [0]{\spacefactor3000\relax}%
\providecommand \BibitemShut  [1]{\csname bibitem#1\endcsname}%
\let\auto@bib@innerbib\@empty
\bibitem [{\citenamefont {He}\ \emph {et~al.}(2018)\citenamefont {He},
  \citenamefont {Zhang}, \citenamefont {Tang}, \citenamefont {Sun},
  \citenamefont {Zhang},\ and\ \citenamefont {Sun}}]{He2018}%
  \BibitemOpen
  \bibfield  {author} {\bibinfo {author} {\bibfnamefont {L.-B.}\ \bibnamefont
  {He}}, \bibinfo {author} {\bibfnamefont {L.}~\bibnamefont {Zhang}}, \bibinfo
  {author} {\bibfnamefont {L.-P.}\ \bibnamefont {Tang}}, \bibinfo {author}
  {\bibfnamefont {J.}~\bibnamefont {Sun}}, \bibinfo {author} {\bibfnamefont
  {Q.-B.}\ \bibnamefont {Zhang}}, \ and\ \bibinfo {author} {\bibfnamefont
  {L.-T.}\ \bibnamefont {Sun}},\ }\bibfield  {title} {\enquote {\bibinfo
  {title} {{Novel behaviors/properties of nanometals induced by surface
  effects}},}\ }\href {\doibase 10.1016/j.mtnano.2018.04.006} {\bibfield
  {journal} {\bibinfo  {journal} {Mater. Today Nano.}\ }\textbf {\bibinfo
  {volume} {1}},\ \bibinfo {pages} {8--21} (\bibinfo {year}
  {2018})}\BibitemShut {NoStop}%
\bibitem [{\citenamefont {Guo}\ \emph {et~al.}(2013)\citenamefont {Guo},
  \citenamefont {Xie},\ and\ \citenamefont {Luo}}]{Guo2013}%
  \BibitemOpen
  \bibfield  {author} {\bibinfo {author} {\bibfnamefont {D.}~\bibnamefont
  {Guo}}, \bibinfo {author} {\bibfnamefont {G.}~\bibnamefont {Xie}}, \ and\
  \bibinfo {author} {\bibfnamefont {J.}~\bibnamefont {Luo}},\ }\bibfield
  {title} {\enquote {\bibinfo {title} {Mechanical properties of nanoparticles:
  basics and applications},}\ }\href {\doibase 10.1088/0022-3727/47/1/013001}
  {\bibfield  {journal} {\bibinfo  {journal} {J. Phys. D: Appl. Phys.}\
  }\textbf {\bibinfo {volume} {47}},\ \bibinfo {pages} {013001} (\bibinfo
  {year} {2013})}\BibitemShut {NoStop}%
\bibitem [{\citenamefont {Buffat}\ and\ \citenamefont
  {Borel}(1976)}]{Buffat1976}%
  \BibitemOpen
  \bibfield  {author} {\bibinfo {author} {\bibfnamefont {Ph.}\ \bibnamefont
  {Buffat}}\ and\ \bibinfo {author} {\bibfnamefont {J-P.}\ \bibnamefont
  {Borel}},\ }\bibfield  {title} {\enquote {\bibinfo {title} {Size effect on
  the melting temperature of gold particles},}\ }\href {\doibase
  10.1103/PhysRevA.13.2287} {\bibfield  {journal} {\bibinfo  {journal} {Phys.
  Rev. A}\ }\textbf {\bibinfo {volume} {13}},\ \bibinfo {pages} {2287--2298}
  (\bibinfo {year} {1976})}\BibitemShut {NoStop}%
\bibitem [{\citenamefont {Amara}\ \emph {et~al.}(2022)\citenamefont {Amara},
  \citenamefont {Nelayah}, \citenamefont {Creuze}, \citenamefont {Chmielewski},
  \citenamefont {Alloyeau}, \citenamefont {Ricolleau},\ and\ \citenamefont
  {Legrand}}]{Amara2022}%
  \BibitemOpen
  \bibfield  {author} {\bibinfo {author} {\bibfnamefont {H.}~\bibnamefont
  {Amara}}, \bibinfo {author} {\bibfnamefont {J.}~\bibnamefont {Nelayah}},
  \bibinfo {author} {\bibfnamefont {J.}~\bibnamefont {Creuze}}, \bibinfo
  {author} {\bibfnamefont {A.}~\bibnamefont {Chmielewski}}, \bibinfo {author}
  {\bibfnamefont {D.}~\bibnamefont {Alloyeau}}, \bibinfo {author}
  {\bibfnamefont {C.}~\bibnamefont {Ricolleau}}, \ and\ \bibinfo {author}
  {\bibfnamefont {B.}~\bibnamefont {Legrand}},\ }\bibfield  {title} {\enquote
  {\bibinfo {title} {Effect of size on the surface energy of noble metal
  nanoparticles from analytical and numerical approaches},}\ }\href {\doibase
  10.1103/PhysRevB.105.165403} {\bibfield  {journal} {\bibinfo  {journal}
  {Phys. Rev. B}\ }\textbf {\bibinfo {volume} {105}},\ \bibinfo {pages}
  {165403} (\bibinfo {year} {2022})}\BibitemShut {NoStop}%
\bibitem [{\citenamefont {Alloyeau}\ \emph {et~al.}(2009)\citenamefont
  {Alloyeau}, \citenamefont {Ricolleau}, \citenamefont {Mottet}, \citenamefont
  {Oikawa}, \citenamefont {Langlois}, \citenamefont {{Le Bouar}}, \citenamefont
  {Braidy},\ and\ \citenamefont {Loiseau}}]{Alloyeau2009}%
  \BibitemOpen
  \bibfield  {author} {\bibinfo {author} {\bibfnamefont {D.}~\bibnamefont
  {Alloyeau}}, \bibinfo {author} {\bibfnamefont {C.}~\bibnamefont {Ricolleau}},
  \bibinfo {author} {\bibfnamefont {C.}~\bibnamefont {Mottet}}, \bibinfo
  {author} {\bibfnamefont {T.}~\bibnamefont {Oikawa}}, \bibinfo {author}
  {\bibfnamefont {C.}~\bibnamefont {Langlois}}, \bibinfo {author}
  {\bibfnamefont {Y.}~\bibnamefont {{Le Bouar}}}, \bibinfo {author}
  {\bibfnamefont {N.}~\bibnamefont {Braidy}}, \ and\ \bibinfo {author}
  {\bibfnamefont {A.}~\bibnamefont {Loiseau}},\ }\bibfield  {title} {\enquote
  {\bibinfo {title} {{Size and shape effects on the order-disorder phase
  transition in CoPt nanoparticles.}}}\ }\href {\doibase 10.1038/nmat2574}
  {\bibfield  {journal} {\bibinfo  {journal} {Nat. Mater.}\ }\textbf {\bibinfo
  {volume} {8}},\ \bibinfo {pages} {940--6} (\bibinfo {year}
  {2009})}\BibitemShut {NoStop}%
\bibitem [{\citenamefont {Uchic}\ \emph {et~al.}(2004)\citenamefont {Uchic},
  \citenamefont {Dimiduk}, \citenamefont {Florando},\ and\ \citenamefont
  {Nix}}]{Uchic2004}%
  \BibitemOpen
  \bibfield  {author} {\bibinfo {author} {\bibfnamefont {M.~D.}\ \bibnamefont
  {Uchic}}, \bibinfo {author} {\bibfnamefont {D.~M.}\ \bibnamefont {Dimiduk}},
  \bibinfo {author} {\bibfnamefont {J.~N.}\ \bibnamefont {Florando}}, \ and\
  \bibinfo {author} {\bibfnamefont {W.~D.}\ \bibnamefont {Nix}},\ }\bibfield
  {title} {\enquote {\bibinfo {title} {Sample dimensions influence strength and
  crystal plasticity},}\ }\href {\doibase 10.1126/science.1098993} {\bibfield
  {journal} {\bibinfo  {journal} {Science}\ }\textbf {\bibinfo {volume}
  {305}},\ \bibinfo {pages} {986--989} (\bibinfo {year} {2004})}\BibitemShut
  {NoStop}%
\bibitem [{\citenamefont {Wu}\ \emph {et~al.}(2005)\citenamefont {Wu},
  \citenamefont {Heidelberg},\ and\ \citenamefont {Boland}}]{Wu2005}%
  \BibitemOpen
  \bibfield  {author} {\bibinfo {author} {\bibfnamefont {B.}~\bibnamefont
  {Wu}}, \bibinfo {author} {\bibfnamefont {A.}~\bibnamefont {Heidelberg}}, \
  and\ \bibinfo {author} {\bibfnamefont {J.~J.}\ \bibnamefont {Boland}},\
  }\bibfield  {title} {\enquote {\bibinfo {title} {{Mechanical properties of
  ultrahigh-strength gold nanowires}},}\ }\href {\doibase 10.1038/nmat1403}
  {\bibfield  {journal} {\bibinfo  {journal} {Nat. Mater.}\ }\textbf {\bibinfo
  {volume} {4}},\ \bibinfo {pages} {525--529} (\bibinfo {year}
  {2005})}\BibitemShut {NoStop}%
\bibitem [{\citenamefont {Kraft}\ \emph {et~al.}(2010)\citenamefont {Kraft},
  \citenamefont {Gruber}, \citenamefont {M\"{o}nig},\ and\ \citenamefont
  {Weygand}}]{Kraft2010}%
  \BibitemOpen
  \bibfield  {author} {\bibinfo {author} {\bibfnamefont {O.}~\bibnamefont
  {Kraft}}, \bibinfo {author} {\bibfnamefont {P.~A.}\ \bibnamefont {Gruber}},
  \bibinfo {author} {\bibfnamefont {R.}~\bibnamefont {M\"{o}nig}}, \ and\
  \bibinfo {author} {\bibfnamefont {D.}~\bibnamefont {Weygand}},\ }\bibfield
  {title} {\enquote {\bibinfo {title} {Plasticity in confined dimensions},}\
  }\href {\doibase 10.1146/annurev-matsci-082908-145409} {\bibfield  {journal}
  {\bibinfo  {journal} {Annu. Rev. Mater. Res.}\ }\textbf {\bibinfo {volume}
  {40}},\ \bibinfo {pages} {293--317} (\bibinfo {year} {2010})}\BibitemShut
  {NoStop}%
\bibitem [{\citenamefont {Deneen}\ \emph {et~al.}(2006)\citenamefont {Deneen},
  \citenamefont {Mook}, \citenamefont {Minor}, \citenamefont {Gerberich},\ and\
  \citenamefont {Carter}}]{Deneen2006}%
  \BibitemOpen
  \bibfield  {author} {\bibinfo {author} {\bibfnamefont {J.}~\bibnamefont
  {Deneen}}, \bibinfo {author} {\bibfnamefont {W.~M.}\ \bibnamefont {Mook}},
  \bibinfo {author} {\bibfnamefont {A.}~\bibnamefont {Minor}}, \bibinfo
  {author} {\bibfnamefont {W.~W.}\ \bibnamefont {Gerberich}}, \ and\ \bibinfo
  {author} {\bibfnamefont {C.~B.}\ \bibnamefont {Carter}},\ }\bibfield  {title}
  {\enquote {\bibinfo {title} {{In situ deformation of silicon nanospheres}},}\
  }\href {\doibase 10.1007/s10853-006-0085-9} {\bibfield  {journal} {\bibinfo
  {journal} {J. Mater. Sci.}\ }\textbf {\bibinfo {volume} {41}},\ \bibinfo
  {pages} {4477--4483} (\bibinfo {year} {2006})}\BibitemShut {NoStop}%
\bibitem [{\citenamefont {Carlton}\ and\ \citenamefont
  {Ferreira}(2012)}]{Carlton2012}%
  \BibitemOpen
  \bibfield  {author} {\bibinfo {author} {\bibfnamefont {C.E.}\ \bibnamefont
  {Carlton}}\ and\ \bibinfo {author} {\bibfnamefont {P.J.}\ \bibnamefont
  {Ferreira}},\ }\bibfield  {title} {\enquote {\bibinfo {title} {In situ tem
  nanoindentation of nanoparticles},}\ }\href {\doibase
  https://doi.org/10.1016/j.micron.2012.03.002} {\bibfield  {journal} {\bibinfo
   {journal} {Micron}\ }\textbf {\bibinfo {volume} {43}},\ \bibinfo {pages}
  {1134--1139} (\bibinfo {year} {2012})}\BibitemShut {NoStop}%
\bibitem [{\citenamefont {Legros}(2014)}]{Legros2014}%
  \BibitemOpen
  \bibfield  {author} {\bibinfo {author} {\bibfnamefont {M.}~\bibnamefont
  {Legros}},\ }\bibfield  {title} {\enquote {\bibinfo {title} {In situ
  mechanical {TEM}: Seeing and measuring under stress with electrons},}\ }\href
  {\doibase https://doi.org/10.1016/j.crhy.2014.02.002} {\bibfield  {journal}
  {\bibinfo  {journal} {C. R. Phys.}\ }\textbf {\bibinfo {volume} {15}},\
  \bibinfo {pages} {224--240} (\bibinfo {year} {2014})}\BibitemShut {NoStop}%
\bibitem [{\citenamefont {de~la Rosa~Abad}\ \emph {et~al.}(2021)\citenamefont
  {de~la Rosa~Abad}, \citenamefont {Londo\~{n}o Calderon}, \citenamefont
  {Bringa}, \citenamefont {Soldano}, \citenamefont {Paz}, \citenamefont
  {Santiago}, \citenamefont {Mej\'{\i}a-Rosales}, \citenamefont {Yacam\'an},\
  and\ \citenamefont {Mariscal}}]{Abad2021}%
  \BibitemOpen
  \bibfield  {author} {\bibinfo {author} {\bibfnamefont {J.~A.}\ \bibnamefont
  {de~la Rosa~Abad}}, \bibinfo {author} {\bibfnamefont {A.}~\bibnamefont
  {Londo\~{n}o Calderon}}, \bibinfo {author} {\bibfnamefont {E.~M.}\
  \bibnamefont {Bringa}}, \bibinfo {author} {\bibfnamefont {G.~J.}\
  \bibnamefont {Soldano}}, \bibinfo {author} {\bibfnamefont {S.~A.}\
  \bibnamefont {Paz}}, \bibinfo {author} {\bibfnamefont {U.}~\bibnamefont
  {Santiago}}, \bibinfo {author} {\bibfnamefont {S.~J.}\ \bibnamefont
  {Mej\'{\i}a-Rosales}}, \bibinfo {author} {\bibfnamefont {M.~J.}\ \bibnamefont
  {Yacam\'an}}, \ and\ \bibinfo {author} {\bibfnamefont {M.~M.}\ \bibnamefont
  {Mariscal}},\ }\bibfield  {title} {\enquote {\bibinfo {title} {Soft or hard?
  investigating the deformation mechanisms of {A}u-{P}d and {P}d nanocubes
  under compression: An experimental and molecular dynamics study},}\ }\href
  {\doibase 10.1021/acs.jpcc.1c07685} {\bibfield  {journal} {\bibinfo
  {journal} {J. Phys. Chem. C}\ }\textbf {\bibinfo {volume} {125}},\ \bibinfo
  {pages} {25298--25306} (\bibinfo {year} {2021})}\BibitemShut {NoStop}%
\bibitem [{\citenamefont {Cuenot}\ \emph {et~al.}(2004)\citenamefont {Cuenot},
  \citenamefont {Fr\'etigny}, \citenamefont {Demoustier-Champagne},\ and\
  \citenamefont {Nysten}}]{Cuenot2004}%
  \BibitemOpen
  \bibfield  {author} {\bibinfo {author} {\bibfnamefont {S.}~\bibnamefont
  {Cuenot}}, \bibinfo {author} {\bibfnamefont {C.}~\bibnamefont {Fr\'etigny}},
  \bibinfo {author} {\bibfnamefont {S.}~\bibnamefont {Demoustier-Champagne}}, \
  and\ \bibinfo {author} {\bibfnamefont {B.}~\bibnamefont {Nysten}},\
  }\bibfield  {title} {\enquote {\bibinfo {title} {Surface tension effect on
  the mechanical properties of nanomaterials measured by atomic force
  microscopy},}\ }\href {\doibase 10.1103/PhysRevB.69.165410} {\bibfield
  {journal} {\bibinfo  {journal} {Phys. Rev. B}\ }\textbf {\bibinfo {volume}
  {69}},\ \bibinfo {pages} {165410} (\bibinfo {year} {2004})}\BibitemShut
  {NoStop}%
\bibitem [{\citenamefont {Shaat}(2019)}]{Shaat2019}%
  \BibitemOpen
  \bibfield  {author} {\bibinfo {author} {\bibfnamefont {M.}~\bibnamefont
  {Shaat}},\ }\bibfield  {title} {\enquote {\bibinfo {title} {Size-dependence
  of young's modulus and poisson's ratio: Effects of material dispersion},}\
  }\href {\doibase https://doi.org/10.1016/j.mechmat.2019.03.012} {\bibfield
  {journal} {\bibinfo  {journal} {Mech. Mater.}\ }\textbf {\bibinfo {volume}
  {133}},\ \bibinfo {pages} {111--119} (\bibinfo {year} {2019})}\BibitemShut
  {NoStop}%
\bibitem [{\citenamefont {Cherian}\ \emph {et~al.}(2010)\citenamefont
  {Cherian}, \citenamefont {Gerard}, \citenamefont {Mahadevan}, \citenamefont
  {Cuong},\ and\ \citenamefont {Maezono}}]{Cherian2010}%
  \BibitemOpen
  \bibfield  {author} {\bibinfo {author} {\bibfnamefont {R.}~\bibnamefont
  {Cherian}}, \bibinfo {author} {\bibfnamefont {C.}~\bibnamefont {Gerard}},
  \bibinfo {author} {\bibfnamefont {P.}~\bibnamefont {Mahadevan}}, \bibinfo
  {author} {\bibfnamefont {Nguyen~Thanh}\ \bibnamefont {Cuong}}, \ and\
  \bibinfo {author} {\bibfnamefont {Ryo}\ \bibnamefont {Maezono}},\ }\bibfield
  {title} {\enquote {\bibinfo {title} {Size dependence of the bulk modulus of
  semiconductor nanocrystals from first-principles calculations},}\ }\href
  {\doibase 10.1103/PhysRevB.82.235321} {\bibfield  {journal} {\bibinfo
  {journal} {Phys. Rev. B}\ }\textbf {\bibinfo {volume} {82}},\ \bibinfo
  {pages} {235321} (\bibinfo {year} {2010})}\BibitemShut {NoStop}%
\bibitem [{\citenamefont {Pizzagalli}(2020)}]{Pizzagalli2020}%
  \BibitemOpen
  \bibfield  {author} {\bibinfo {author} {\bibfnamefont {L.}~\bibnamefont
  {Pizzagalli}},\ }\bibfield  {title} {\enquote {\bibinfo {title}
  {Finite-temperature mechanical properties of nanostructures with
  first-principles accuracy},}\ }\href {\doibase 10.1103/PhysRevB.102.094102}
  {\bibfield  {journal} {\bibinfo  {journal} {Phys. Rev. B}\ }\textbf {\bibinfo
  {volume} {102}},\ \bibinfo {pages} {094102} (\bibinfo {year}
  {2020})}\BibitemShut {NoStop}%
\bibitem [{\citenamefont {Amodeo}\ and\ \citenamefont
  {Pizzagalli}(2021)}]{Amodeo2021}%
  \BibitemOpen
  \bibfield  {author} {\bibinfo {author} {\bibfnamefont {J.}~\bibnamefont
  {Amodeo}}\ and\ \bibinfo {author} {\bibfnamefont {L.}~\bibnamefont
  {Pizzagalli}},\ }\bibfield  {title} {\enquote {\bibinfo {title} {Modeling the
  mechanical properties of nanoparticles: a review},}\ }\href {\doibase
  10.5802/crphys.70} {\bibfield  {journal} {\bibinfo  {journal} {C. R. Phys.}\
  }\textbf {\bibinfo {volume} {22}},\ \bibinfo {pages} {35--66} (\bibinfo
  {year} {2021})}\BibitemShut {NoStop}%
\bibitem [{\citenamefont {Mordehai}\ \emph {et~al.}(2011)\citenamefont
  {Mordehai}, \citenamefont {Lee}, \citenamefont {Backes}, \citenamefont
  {Srolovitz}, \citenamefont {Nix},\ and\ \citenamefont
  {Rabkin}}]{Mordehai2011}%
  \BibitemOpen
  \bibfield  {author} {\bibinfo {author} {\bibfnamefont {D.}~\bibnamefont
  {Mordehai}}, \bibinfo {author} {\bibfnamefont {S.~W.}\ \bibnamefont {Lee}},
  \bibinfo {author} {\bibfnamefont {B.}~\bibnamefont {Backes}}, \bibinfo
  {author} {\bibfnamefont {D.~J.}\ \bibnamefont {Srolovitz}}, \bibinfo {author}
  {\bibfnamefont {W.~D.}\ \bibnamefont {Nix}}, \ and\ \bibinfo {author}
  {\bibfnamefont {E.}~\bibnamefont {Rabkin}},\ }\bibfield  {title} {\enquote
  {\bibinfo {title} {{Size effect in compression of single-crystal gold
  microparticles}},}\ }\href {\doibase 10.1016/j.actamat.2011.04.057}
  {\bibfield  {journal} {\bibinfo  {journal} {Acta Mater.}\ }\textbf {\bibinfo
  {volume} {59}},\ \bibinfo {pages} {5202--5215} (\bibinfo {year}
  {2011})}\BibitemShut {NoStop}%
\bibitem [{\citenamefont {Feruz}\ and\ \citenamefont
  {Mordehai}(2016)}]{Feruz2016}%
  \BibitemOpen
  \bibfield  {author} {\bibinfo {author} {\bibfnamefont {Y.}~\bibnamefont
  {Feruz}}\ and\ \bibinfo {author} {\bibfnamefont {D.}~\bibnamefont
  {Mordehai}},\ }\bibfield  {title} {\enquote {\bibinfo {title} {{Towards a
  universal size-dependent strength of face-centered cubic nanoparticles}},}\
  }\href {\doibase 10.1016/j.actamat.2015.10.027} {\bibfield  {journal}
  {\bibinfo  {journal} {Acta Mater.}\ }\textbf {\bibinfo {volume} {103}},\
  \bibinfo {pages} {433--441} (\bibinfo {year} {2016})}\BibitemShut {NoStop}%
\bibitem [{\citenamefont {Kilymis}\ \emph {et~al.}(2018)\citenamefont
  {Kilymis}, \citenamefont {G{\'{e}}rard}, \citenamefont {Amodeo},
  \citenamefont {Waghmare},\ and\ \citenamefont {Pizzagalli}}]{Kilymis2018}%
  \BibitemOpen
  \bibfield  {author} {\bibinfo {author} {\bibfnamefont {D.}~\bibnamefont
  {Kilymis}}, \bibinfo {author} {\bibfnamefont {C.}~\bibnamefont
  {G{\'{e}}rard}}, \bibinfo {author} {\bibfnamefont {J.}~\bibnamefont
  {Amodeo}}, \bibinfo {author} {\bibfnamefont {U.~V.}\ \bibnamefont
  {Waghmare}}, \ and\ \bibinfo {author} {\bibfnamefont {L.}~\bibnamefont
  {Pizzagalli}},\ }\bibfield  {title} {\enquote {\bibinfo {title} {{Uniaxial
  compression of silicon nanoparticles: An atomistic study on the shape and
  size effects}},}\ }\href {\doibase 10.1016/j.actamat.2018.07.063} {\bibfield
  {journal} {\bibinfo  {journal} {Acta Mater.}\ }\textbf {\bibinfo {volume}
  {158}},\ \bibinfo {pages} {155--166} (\bibinfo {year} {2018})}\BibitemShut
  {NoStop}%
\bibitem [{\citenamefont {de~la Rosa~Abad}\ \emph {et~al.}(2023)\citenamefont
  {de~la Rosa~Abad}, \citenamefont {Bringa}, \citenamefont
  {Mej\'{\i}a-Rosales},\ and\ \citenamefont {Mariscal}}]{Abad2023}%
  \BibitemOpen
  \bibfield  {author} {\bibinfo {author} {\bibfnamefont {J.~A.}\ \bibnamefont
  {de~la Rosa~Abad}}, \bibinfo {author} {\bibfnamefont {E.~M.}\ \bibnamefont
  {Bringa}}, \bibinfo {author} {\bibfnamefont {S.~J.}\ \bibnamefont
  {Mej\'{\i}a-Rosales}}, \ and\ \bibinfo {author} {\bibfnamefont {M.~M.}\
  \bibnamefont {Mariscal}},\ }\bibfield  {title} {\enquote {\bibinfo {title}
  {On the mechanical response in nanoalloys: the case of {N}i{C}o},}\ }\href
  {\doibase 10.1039/D2FD00111J} {\bibfield  {journal} {\bibinfo  {journal}
  {Faraday Discuss.}\ }\textbf {\bibinfo {volume} {242}},\ \bibinfo {pages}
  {23--34} (\bibinfo {year} {2023})}\BibitemShut {NoStop}%
\bibitem [{\citenamefont {Henry}(2005)}]{Henry2005}%
  \BibitemOpen
  \bibfield  {author} {\bibinfo {author} {\bibfnamefont {C.~R.}\ \bibnamefont
  {Henry}},\ }\bibfield  {title} {\enquote {\bibinfo {title} {{Morphology of
  supported nanoparticles}},}\ }\href {\doibase 10.1016/j.progsurf.2005.09.004}
  {\bibfield  {journal} {\bibinfo  {journal} {Progr. Surf. Sci.}\ }\textbf
  {\bibinfo {volume} {80}},\ \bibinfo {pages} {92--116} (\bibinfo {year}
  {2005})}\BibitemShut {NoStop}%
\bibitem [{\citenamefont {Wulff}(1901)}]{wulff1901}%
  \BibitemOpen
  \bibfield  {author} {\bibinfo {author} {\bibfnamefont {G.}~\bibnamefont
  {Wulff}},\ }\bibfield  {title} {\enquote {\bibinfo {title} {{Zur Frage der
  Geschwindigkeit des Wachsthums und der Aufl\"osung der Krystallfl\"achen}},}\
  }\href {\doibase https://doi.org/10.1524/zkri.1901.34.1.449} {\bibfield
  {journal} {\bibinfo  {journal} {Z. Kristallogr.}\ }\textbf {\bibinfo {volume}
  {34}},\ \bibinfo {pages} {449} (\bibinfo {year} {1901})}\BibitemShut
  {NoStop}%
\bibitem [{\citenamefont {Kaischew}(1952)}]{Kaischew1952}%
  \BibitemOpen
  \bibfield  {author} {\bibinfo {author} {\bibfnamefont {R.}~\bibnamefont
  {Kaischew}},\ }\href@noop {} {\emph {\bibinfo {title} {Arbeitstagung
  Festk\"orper Physik}}}\ (\bibinfo  {publisher} {Dresden},\ \bibinfo {year}
  {1952})\ p.~\bibinfo {pages} {81}\BibitemShut {NoStop}%
\bibitem [{\citenamefont {Plimpton}(1995)}]{Plimpton1995}%
  \BibitemOpen
  \bibfield  {author} {\bibinfo {author} {\bibfnamefont {S.}~\bibnamefont
  {Plimpton}},\ }\bibfield  {title} {\enquote {\bibinfo {title} {Fast parallel
  algorithms for short-range molecular dynamics},}\ }\href {\doibase
  https://doi.org/10.1006/jcph.1995.1039} {\bibfield  {journal} {\bibinfo
  {journal} {J. Comput. Phys.}\ }\textbf {\bibinfo {volume} {117}},\ \bibinfo
  {pages} {1--19} (\bibinfo {year} {1995})}\BibitemShut {NoStop}%
\bibitem [{\citenamefont {Roy}\ \emph {et~al.}(2019)\citenamefont {Roy},
  \citenamefont {Gatti}, \citenamefont {Devincre},\ and\ \citenamefont
  {Mordehai}}]{Roy2019}%
  \BibitemOpen
  \bibfield  {author} {\bibinfo {author} {\bibfnamefont {S.}~\bibnamefont
  {Roy}}, \bibinfo {author} {\bibfnamefont {R.}~\bibnamefont {Gatti}}, \bibinfo
  {author} {\bibfnamefont {B.}~\bibnamefont {Devincre}}, \ and\ \bibinfo
  {author} {\bibfnamefont {D.}~\bibnamefont {Mordehai}},\ }\bibfield  {title}
  {\enquote {\bibinfo {title} {{A multiscale study of the size-effect in
  nanoindentation of Au nanoparticles}},}\ }\href {\doibase
  10.1016/j.commatsci.2019.02.013} {\bibfield  {journal} {\bibinfo  {journal}
  {Comput. Mater. Sci.}\ }\textbf {\bibinfo {volume} {162}},\ \bibinfo {pages}
  {47--59} (\bibinfo {year} {2019})}\BibitemShut {NoStop}%
\bibitem [{\citenamefont {Ducastelle}(1970)}]{Ducastelle1970}%
  \BibitemOpen
  \bibfield  {author} {\bibinfo {author} {\bibfnamefont {F.}~\bibnamefont
  {Ducastelle}},\ }\bibfield  {title} {\enquote {\bibinfo {title} {{Module
  \'elastique des m\'etaux de transition}},}\ }\href {\doibase
  https://doi.org/10.1051/jphys:019700031011-120105500} {\bibfield  {journal}
  {\bibinfo  {journal} {J. Phys. (Paris)}\ }\textbf {\bibinfo {volume} {31}},\
  \bibinfo {pages} {1055} (\bibinfo {year} {1970})}\BibitemShut {NoStop}%
\bibitem [{\citenamefont {Rosato}\ \emph {et~al.}(1989)\citenamefont {Rosato},
  \citenamefont {Guillop\'e},\ and\ \citenamefont {Legrand}}]{Rosato1989}%
  \BibitemOpen
  \bibfield  {author} {\bibinfo {author} {\bibfnamefont {V.}~\bibnamefont
  {Rosato}}, \bibinfo {author} {\bibfnamefont {M.}~\bibnamefont {Guillop\'e}},
  \ and\ \bibinfo {author} {\bibfnamefont {B.}~\bibnamefont {Legrand}},\
  }\bibfield  {title} {\enquote {\bibinfo {title} {{Thermodynamical and
  structural properties of f.c.c. transition metals using a simple
  tight-binding model}},}\ }\href {\doibase DOI: 10.1080/01418618908205062}
  {\bibfield  {journal} {\bibinfo  {journal} {Philos. Mag. A}\ }\textbf
  {\bibinfo {volume} {59}},\ \bibinfo {pages} {321} (\bibinfo {year}
  {1989})}\BibitemShut {NoStop}%
\bibitem [{\citenamefont {Kittel}(1995)}]{Kittel1995}%
  \BibitemOpen
  \bibfield  {author} {\bibinfo {author} {\bibfnamefont {C.}~\bibnamefont
  {Kittel}},\ }\href@noop {} {\emph {\bibinfo {title} {Introduction to Solid
  State Physics}}}\ (\bibinfo  {publisher} {Wiley, New York},\ \bibinfo {year}
  {1995})\BibitemShut {NoStop}%
\bibitem [{\citenamefont {Simmons}\ and\ \citenamefont
  {Wang}(1971)}]{Simmons1971}%
  \BibitemOpen
  \bibfield  {author} {\bibinfo {author} {\bibfnamefont {G.}~\bibnamefont
  {Simmons}}\ and\ \bibinfo {author} {\bibfnamefont {H.}~\bibnamefont {Wang}},\
  }\href@noop {} {\emph {\bibinfo {title} {Single Crystal Elastic Constants and
  Calculated Aggregates Properties}}}\ (\bibinfo  {publisher} {MIT,
  Cambridge},\ \bibinfo {year} {1971})\BibitemShut {NoStop}%
\bibitem [{\citenamefont {Delfour}\ \emph {et~al.}(2009)\citenamefont
  {Delfour}, \citenamefont {Creuze},\ and\ \citenamefont
  {Legrand}}]{Delfour2009}%
  \BibitemOpen
  \bibfield  {author} {\bibinfo {author} {\bibfnamefont {L.}~\bibnamefont
  {Delfour}}, \bibinfo {author} {\bibfnamefont {J.}~\bibnamefont {Creuze}}, \
  and\ \bibinfo {author} {\bibfnamefont {B.}~\bibnamefont {Legrand}},\
  }\bibfield  {title} {\enquote {\bibinfo {title} {Exotic behavior of the outer
  shell of bimetallic nanoalloys},}\ }\href {\doibase
  10.1103/PhysRevLett.103.205701} {\bibfield  {journal} {\bibinfo  {journal}
  {Phys. Rev. Lett.}\ }\textbf {\bibinfo {volume} {103}},\ \bibinfo {pages}
  {205701} (\bibinfo {year} {2009})}\BibitemShut {NoStop}%
\bibitem [{\citenamefont {Chmielewski}\ \emph {et~al.}(2018)\citenamefont
  {Chmielewski}, \citenamefont {Nelayah}, \citenamefont {Amara}, \citenamefont
  {Creuze}, \citenamefont {Alloyeau}, \citenamefont {Wang},\ and\ \citenamefont
  {Ricolleau}}]{Chmielewski2018}%
  \BibitemOpen
  \bibfield  {author} {\bibinfo {author} {\bibfnamefont {A.}~\bibnamefont
  {Chmielewski}}, \bibinfo {author} {\bibfnamefont {J.}~\bibnamefont
  {Nelayah}}, \bibinfo {author} {\bibfnamefont {H.}~\bibnamefont {Amara}},
  \bibinfo {author} {\bibfnamefont {J.}~\bibnamefont {Creuze}}, \bibinfo
  {author} {\bibfnamefont {D.}~\bibnamefont {Alloyeau}}, \bibinfo {author}
  {\bibfnamefont {G.}~\bibnamefont {Wang}}, \ and\ \bibinfo {author}
  {\bibfnamefont {C.}~\bibnamefont {Ricolleau}},\ }\bibfield  {title} {\enquote
  {\bibinfo {title} {{Direct Measurement of the Surface Energy of Bimetallic
  Nanoparticles: Evidence of Vegard's Rulelike Dependence}},}\ }\href {\doibase
  10.1103/PhysRevLett.120.025901} {\bibfield  {journal} {\bibinfo  {journal}
  {Phys. Rev. Lett.}\ }\textbf {\bibinfo {volume} {120}},\ \bibinfo {pages}
  {025901} (\bibinfo {year} {2018})}\BibitemShut {NoStop}%
\bibitem [{\citenamefont {Front}\ and\ \citenamefont
  {Mottet}(2021)}]{Front2021}%
  \BibitemOpen
  \bibfield  {author} {\bibinfo {author} {\bibfnamefont {A.}~\bibnamefont
  {Front}}\ and\ \bibinfo {author} {\bibfnamefont {C.}~\bibnamefont {Mottet}},\
  }\bibfield  {title} {\enquote {\bibinfo {title} {Stress effect on segregation
  and ordering in {P}t-{A}g nanoalloys},}\ }\href {\doibase
  10.1088/1361-648x/abe07a} {\bibfield  {journal} {\bibinfo  {journal} {J.
  Phys.: Condens. Matter}\ }\textbf {\bibinfo {volume} {33}},\ \bibinfo {pages}
  {154006} (\bibinfo {year} {2021})}\BibitemShut {NoStop}%
\bibitem [{\citenamefont {Aln{\ae}s}\ \emph {et~al.}(2015)\citenamefont
  {Aln{\ae}s}, \citenamefont {Blechta}, \citenamefont {Hake}, \citenamefont
  {Johansson}, \citenamefont {Kehlet}, \citenamefont {Logg}, \citenamefont
  {Richardson}, \citenamefont {Ring}, \citenamefont {Rognes},\ and\
  \citenamefont {Wells}}]{Fenics}%
  \BibitemOpen
  \bibfield  {author} {\bibinfo {author} {\bibfnamefont {M.~S.}\ \bibnamefont
  {Aln{\ae}s}}, \bibinfo {author} {\bibfnamefont {J.}~\bibnamefont {Blechta}},
  \bibinfo {author} {\bibfnamefont {J.}~\bibnamefont {Hake}}, \bibinfo {author}
  {\bibfnamefont {A.}~\bibnamefont {Johansson}}, \bibinfo {author}
  {\bibfnamefont {B.}~\bibnamefont {Kehlet}}, \bibinfo {author} {\bibfnamefont
  {A.}~\bibnamefont {Logg}}, \bibinfo {author} {\bibfnamefont {C.}~\bibnamefont
  {Richardson}}, \bibinfo {author} {\bibfnamefont {J.}~\bibnamefont {Ring}},
  \bibinfo {author} {\bibfnamefont {M.~E.}\ \bibnamefont {Rognes}}, \ and\
  \bibinfo {author} {\bibfnamefont {G.~N.}\ \bibnamefont {Wells}},\ }\bibfield
  {title} {\enquote {\bibinfo {title} {The {FEniCS} {P}roject {V}ersion 1.5},}\
  }\href {\doibase 10.11588/ans.2015.100.20553} {\bibfield  {journal} {\bibinfo
   {journal} {Archive of Numerical Software}\ }\textbf {\bibinfo {volume}
  {3}},\ \bibinfo {pages} {9--23} (\bibinfo {year} {2015})}\BibitemShut
  {NoStop}%
\bibitem [{\citenamefont {Jacobs}\ and\ \citenamefont
  {Martini}(2017)}]{Jacobs2017}%
  \BibitemOpen
  \bibfield  {author} {\bibinfo {author} {\bibfnamefont {T.~D.~B.}\
  \bibnamefont {Jacobs}}\ and\ \bibinfo {author} {\bibfnamefont
  {A.}~\bibnamefont {Martini}},\ }\bibfield  {title} {\enquote {\bibinfo
  {title} {{Measuring and Understanding Contact Area at the Nanoscale: A
  Review}},}\ }\href {\doibase 10.1115/1.4038130} {\bibfield  {journal}
  {\bibinfo  {journal} {Appl. Mech. Rev.}\ }\textbf {\bibinfo {volume} {69}},\
  \bibinfo {pages} {060802} (\bibinfo {year} {2017})}\BibitemShut {NoStop}%
\bibitem [{\citenamefont {Amodeo}\ and\ \citenamefont
  {Lizoul}(2017)}]{Amodeo2017}%
  \BibitemOpen
  \bibfield  {author} {\bibinfo {author} {\bibfnamefont {J.}~\bibnamefont
  {Amodeo}}\ and\ \bibinfo {author} {\bibfnamefont {K.}~\bibnamefont
  {Lizoul}},\ }\bibfield  {title} {\enquote {\bibinfo {title} {Mechanical
  properties and dislocation nucleation in nanocrystals with blunt edges},}\
  }\href {\doibase https://doi.org/10.1016/j.matdes.2017.09.009} {\bibfield
  {journal} {\bibinfo  {journal} {Mater. Des.}\ }\textbf {\bibinfo {volume}
  {135}},\ \bibinfo {pages} {223--231} (\bibinfo {year} {2017})}\BibitemShut
  {NoStop}%
\bibitem [{\citenamefont {Armstrong}\ and\ \citenamefont
  {Peukert}(2012)}]{Armstrong2012}%
  \BibitemOpen
  \bibfield  {author} {\bibinfo {author} {\bibfnamefont {P.}~\bibnamefont
  {Armstrong}}\ and\ \bibinfo {author} {\bibfnamefont {W.}~\bibnamefont
  {Peukert}},\ }\bibfield  {title} {\enquote {\bibinfo {title} {Size effects in
  the elastic deformation behavior of metallic nanoparticles},}\ }\href
  {\doibase 10.1007/s11051-012-1288-4} {\bibfield  {journal} {\bibinfo
  {journal} {J. Nanopart. Res.}\ }\textbf {\bibinfo {volume} {14}} (\bibinfo
  {year} {2012}),\ 10.1007/s11051-012-1288-4}\BibitemShut {NoStop}%
\bibitem [{\citenamefont {Gupta}(1981)}]{Gupta1981}%
  \BibitemOpen
  \bibfield  {author} {\bibinfo {author} {\bibfnamefont {R.~P.}\ \bibnamefont
  {Gupta}},\ }\bibfield  {title} {\enquote {\bibinfo {title} {Lattice
  relaxation at a metal surface},}\ }\href {\doibase 10.1103/PhysRevB.23.6265}
  {\bibfield  {journal} {\bibinfo  {journal} {Phys. Rev. B}\ }\textbf {\bibinfo
  {volume} {23}},\ \bibinfo {pages} {6265--6270} (\bibinfo {year}
  {1981})}\BibitemShut {NoStop}%
\end{thebibliography}
\end{document}